\documentclass[letterpaper,english,iop]{emulateapj}
\usepackage[LGR,T1]{fontenc}
\usepackage{color}
\usepackage{babel}
\usepackage{array}
\usepackage{booktabs}
\usepackage{textcomp}
\usepackage{multirow}
\usepackage{graphicx}
\usepackage{wasysym}
\usepackage[unicode=true]
 {hyperref}
\usepackage{breakurl}

\makeatletter

\ProvideTextCommand{\~}{LGR}[1]{\char126#1}

\shorttitle{Highly ionized gas in the Milky Way}
\shortauthors{JO ET AL.}

\newenvironment{continuedfigure}{%
\addtocounter{figure}{-1}%
\begin{figure}}{%
\end{figure}}

\newenvironment{continuedfigure*}{%
\addtocounter{figure}{-1}%
\begin{figure}}{%
\end{figure}}

\makeatother

\begin{document}

\title{Global distribution of far-ultraviolet emissions from highly ionized gas in the Milky Way}

\author{Young-Soo Jo\altaffilmark{1}, Kwang-il Seon\altaffilmark{1,2}, Kyoung-Wook Min\altaffilmark{3}, Jerry Edelstein\altaffilmark{4}, and Wonyong Han\altaffilmark{1}}

\altaffiltext{1}{Korea Astronomy and Space Science Institute (KASI), 776 Daedeokdae-ro, Yuseong-gu, Daejeon, 34055, Korea; stspeak@gmail.com}
\altaffiltext{2}{Astronomy and Space Science Major, University of Science and Technology (UST), Korea, 217 Gajeong-ro, Yuseong-gu, Daejeon, 34113, Korea}
\altaffiltext{3}{Korea Advanced Institute of Science and Technology (KAIST), 291, Daehak-ro, Yuseong-gu, Daejeon, 34141, Korea}
\altaffiltext{4}{Space Sciences Laboratory, University of California, Berkeley, CA 94720, USA}

\begin{abstract}
We present all-sky maps of two major FUV cooling lines, \ion{C}{4} and \ion{O}{6}, of highly ionized gas to investigate the nature of the transition-temperature gas. From the extinction-corrected line intensities of \ion{C}{4} and \ion{O}{6}, we calculated the gas temperature and the emission measure of the transition-temperature gas assuming isothermal plasma in the collisional ionization equilibrium. The gas temperature was found to be more or less uniform throughout the Galaxy with a value of (1.89 $\pm$ 0.06) $\times$ $10^5$ K. The emission measure of the transition-temperature gas is described well by a disk-like model in which the scale height of the electron density is $z_0=6_{-2}^{+3}$ kpc. The total mass of the transition-temperature gas is estimated to be approximately $6.4_{-2.8}^{+5.2}\times10^9 M_{\astrosun}$. We also calculated the volume-filling fraction of the transition-temperature gas, which was estimated to be $f=0.26\pm0.09$, and varies from $f\sim0.37$ in the inner Galaxy to $f\sim0.18$ in the outer Galaxy. The spatial distribution of \ion{C}{4} and \ion{O}{6} cannot be explained by a simple supernova remnant model or a three-phase model. The combined effects of supernova remnants and turbulent mixing layers can explain the intensity ratio of \ion{C}{4} and \ion{O}{6}. Thermal conduction front models and high-velocity cloud models are also consistent with our observation.
\end{abstract}


\section{INTRODUCTION}

It is well known that the interstellar medium (ISM) plays a key role in the evolutionary cycle of matter in our Galaxy. Further, knowledge on its distribution as well as its physical and chemical properties is essential to understanding the formation and evolution of stars, planets, and even the Milky Way Galaxy itself. ISM has been conveniently classified into a few approximately stable phases according to its temperature and density, although its detailed properties may vary significantly within each phase. \citet{Fie1969} suggested a two-phase model of the cold neutral medium (CNM) with a temperature of $\sim$$10^2$ K and the warm neutral medium (WNM) with a temperature of $\sim$$10^4$ K, which are in pressure equilibrium with each other. \citet{McK1977} added another phase of temperature of $\sim$$10^6$ K, known as the hot ionized medium (HIM), as its cooling time is much longer than that of the transition-temperature gas, whose temperature is in the range of $10^4 < T < 10^6$ K. Finally, the warm ionized medium (WIM), the density and temperature of which are comparable to those of the WNM, has been observed ubiquitously as a major phase of the ionized hydrogen of ISM \citep[e.g.,][]{Haf2009} \citep[cf.,][]{Seo2012}. Whereas, these simple classifications may be useful for understanding the basic morphologies of ISM, the actual ISM is rarely in a stable state and, instead, it may show complex and even turbulent motions.

There is ample evidence that the Galactic halo is filled with a diffuse hot gas. Approximately 60 years ago, \citet{Spi1956} predicted the existence of the background hot gas far above the Galactic disk based on the observations of cold clouds, as these clouds should be in dynamical equilibrium with the surrounding hot gas. The existence of this diffuse hot gas was finally confirmed roughly 10 years later by direct observations of the soft X-ray background \citep{Bow1968}. Since then, many improved X-ray observations have significantly advanced our knowledge on the physical and morphological properties of the Milky Way hot gas \citep{McC1983,Mar1984,Sno1997}. A common consensus on the generation of the hot gas is that supernova explosions play a major role; however, we still have an insufficient understanding of how this diffuse hot gas forms and evolves in the Galactic scale. There have been several representative theoretical models, including an isolated supernova-type model \citep{Cox1974}, a three-phase model \citep{McK1977}, and a Galactic fountain model \citep{Sha1976}. In the isolated supernova-type model, \citet{Cox1974} argued that a supernova rate of one per 50 years is enough to maintain hot gas percolated in the ISM by considering the evolution of individual supernova remnants as well as the probability of overlap of these remnants. \citet{McK1977} recalculated the probability of overlap of supernova remnants and found it to be close to unity, which implies that hot gas is not easily dissipated and is prevalent in ISM. They went on to propose a three-phase model, in which warm/cold clouds are surrounded by hot gases generated by supernovae. \citet{Sha1976} pointed out the inefficiency of the cooling of hot gas by conduction or radiation in the disk and proposed a galactic fountain model in which convection away from the disk, and the subsequent radiation, may effectively cool the hot gas. According to this model, the hot gas cools and condenses to form clouds above $\sim$$1$ kpc from the disk, that fall back to the disk at high velocities. In addition to these traditional models, many advanced and hybrid models have also been proposed \citep{Edg1986, Ito1988, Bor1990, Hou1990, Sla1993a, Shu1994, She2018}. Recently, hydrodynamic simulations on a Galactic scale have been conducted to predict the spatial distribution of the multiphase ISM \citep{Hil2012, Vas2015, Kim2017, Kim2018}. However, none of these models fully reproduce the observations, the predictions of the global characteristics of ISM vary significantly according to the models, as well as the parameters adopted for the models such as supernova explosion rates. For instance, the volume-filling fraction of the hot gas is found to vary from 10\% to 95\% \citep{Cox1974, McK1977, Sla1993a}. More recently, \citet{deA2004} showed that the volume-filling fraction can vary from approximately 17\% for the Galactic SN rate to 44\% for the 16-times Galactic SN rate. Hence, it is evident that observations of ISM of various phases on a global scale, including the transition-temperature gas phase, are crucial to validate the models, including their parameters.

One of the keys to understanding the hot gas evolution on a global scale is to determine the global spatial distribution and volume-filling fraction of the transition-temperature gas in our Galaxy, as a transition-temperature gas of $10^4 < T < 10^6$ K is generated when a hot gas ($\sim$$10^6$ K) interacts with a colder gas (T < $10^4$ K). Hence, this short-lived gas could be direct evidence of interaction between the HIM and WIM/WNM. Prominent emission lines associated with the cooling gas are in the far-ultraviolet (FUV) wavelengths: \ion{O}{6} ($\lambda\lambda$1032,1038 \AA), \ion{Si}{4} ($\lambda\lambda$1394,1403 \AA), and \ion{C}{4} ($\lambda\lambda$1548,1551 \AA) are such examples. These lines may serve as effective tracers for the transition-temperature gas of $10^{4.5}$\textendash$10^{5.5}$ K. Observations of these FUV lines have been performed mainly by the absorption line measurements of major missions including Berkeley Far-UV Shuttle Telescope \citep{Mar1990}, International Ultraviolet Explorer \citep{Sem1992}, Hubble Space Telescope \citep{Sav1993}, and Far Ultraviolet Spectroscopic Explorer \citep[FUSE;][]{Moo2000, Sav2000}. As absorption line observations can only provide information on limited directions along the line of sight, only a rough estimation can be made on the spatial distribution of the transition-temperature gas with a large uncertainty. The scale height of the transition-temperature gas was found to be as large as several kilo parsecs \citep{Sav1993, Sav1997, Sav2000, Sav2003, Bow2008, Sav2009}.

Therefore, it is obvious that observation of the FUV emission lines for the entire sky of the Milky Way is essential for understanding the global scale distribution of the transition-temperature gas \citep{Shu1994, Kor1998}. In this regard, a spectral survey in the FUV wavelengths made by the Far-Ultraviolet Imaging Spectrograph (FIMS), also known as Spectroscopy of Plasma Evolution from Astrophysical Radiation \citep[SPEAR;][]{Ede2006a, Ede2006b}, could be a valuable source of information on the Galactic morphology of the transition-temperature gas. In the present study, we report all-sky maps of the two major FUV cooling lines, \ion{C}{4} and \ion{O}{6}, constructed from the archival data of FIMS/SPEAR observations. Section 2 describes the procedure of construction of the emission line maps. In Section 3.1, we describe the Galactic morphology and physical properties of the transition-temperature gas traced by \ion{C}{4} and \ion{O}{6}. The volume-filling fraction of the transition-temperature gas is discussed in Section 3.2. We also discuss the validation of the hot gas generation models in Section 4. Finally, a summary of the study is given in Section 5.

\section{Data reduction}

\subsection{Construction of a Three-dimensional S-band Data Cube}

FIMS/SPEAR is a dual-channel imaging spectrograph comprising a short-wavelength channel (\textit{S}-band; 900—1150 \AA{} with a spectral resolution of $\lambda/\Delta\lambda\sim550$ and 4\degr.0$\times$4\arcmin.6 field of view) and a long-wavelength channel (\textit{L}-band; 1350—1750 \AA{} with a spectral resolution of $\lambda/\Delta\lambda\sim550$ and 7\degr.4$\times$4\arcmin.3 field of view). It is the main payload of STSAT-1, the first Korean scientific satellite, launched on September 27, 2003 and operated for more than a year and a half in the sun-synchronous orbit of 98\degr.2 with the equatorial crossing of 22:00 local time at an altitude of $\sim$685 km. FIMS/SPEAR obtained the survey data of $\sim$1,500 orbits during its mission lifetime, with a coverage of $\sim$86\% of the sky. However, the original data provided inaccurate coordinate information for some datasets due to the attitude control system problem. We recently constructed a three-dimensional (3-D) data cube of \textit{L}-band data with a pixel size of $\sim$0.92\degr{} and a spectral bin of 1 \AA, through manual attitude correction and the elaborate elimination of instrumental noise. Details of the \textit{L}-band data cube are described in \citet{Jo2017}, in which the distribution of Galactic molecular hydrogen is studied based on this \textit{L}-band data cube. In a similar manner, we also constructed a 3-D data cube of the \textit{S}-band data.

\begin{figure}[t]
	\begin{centering}
		\medskip{}
		\par\end{centering}
	\begin{centering}
		\includegraphics[clip,scale=0.6]{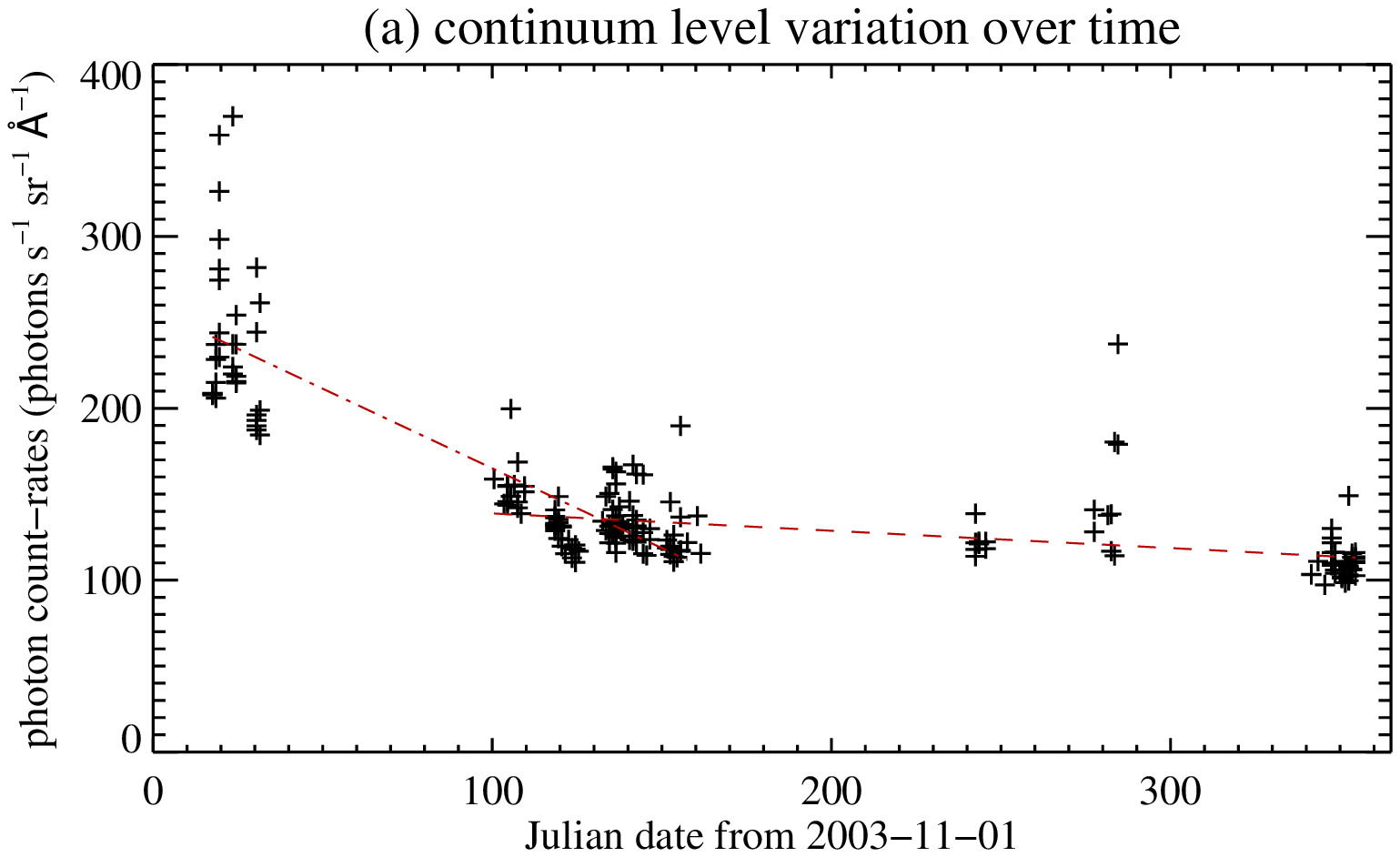}
		\par\end{centering}
	\begin{centering}
		\medskip{}
		\par\end{centering}
	\begin{centering}
		\includegraphics[clip,scale=0.6]{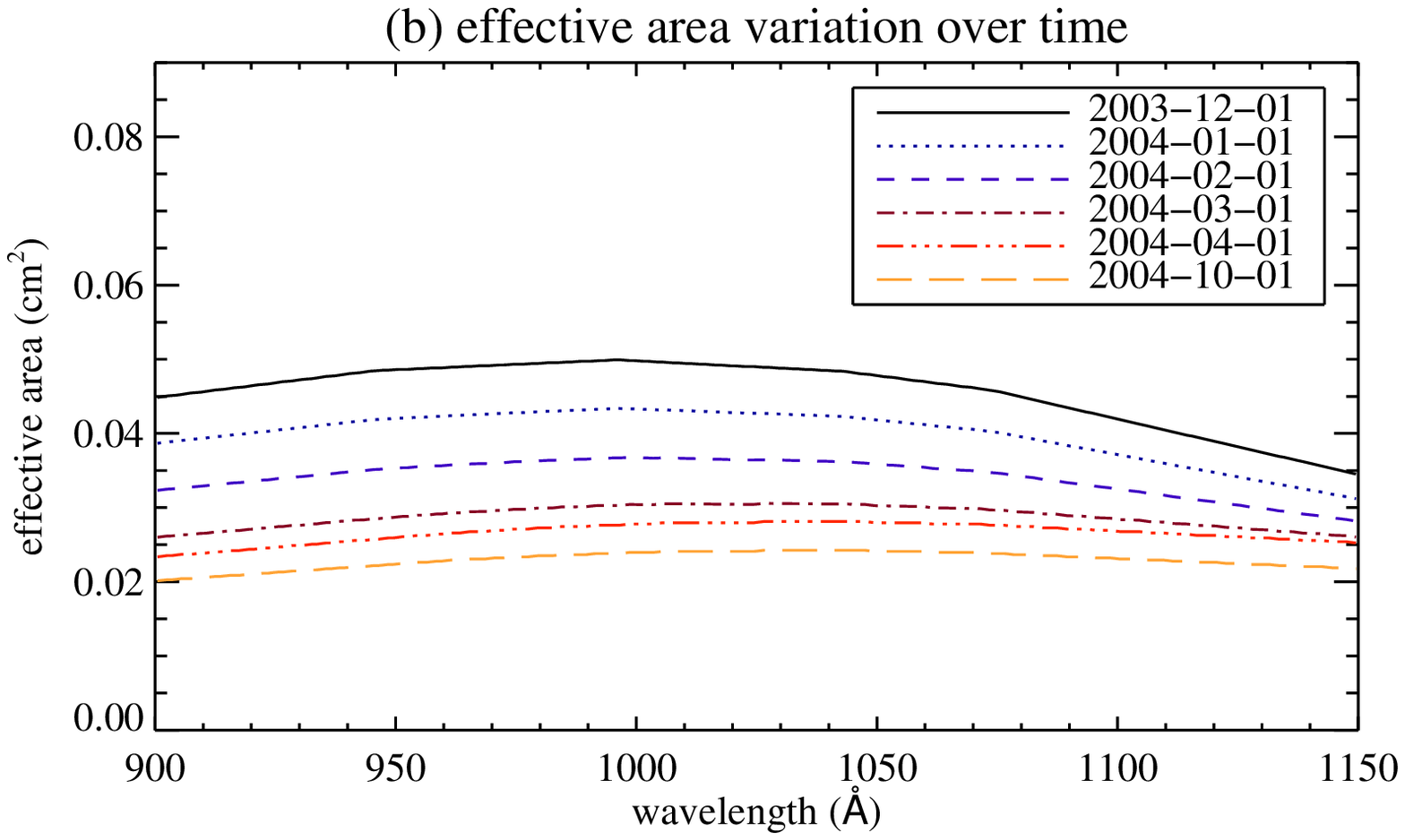}
		\par\end{centering}
	\begin{centering}
		\medskip{}
		\par\end{centering}
	\caption{\label{fig1}(a) Variation of continuum level of \textit{S}-band over time, and (b) Variation of effective area over time.}
\end{figure}

To construct a 3-D data cube for the \textit{S}-band, we started with a photon count-rate map based on the HEALPix scheme \citep{Gor2005} of a resolution parameter $N_{side}$ = 512 (a pixel size of $\sim$6.9 arcmin), as in the case of the \textit{L}-band. The bright point sources were removed in the same way as for the \textit{L}-band, with the aid of two-dimensional Gaussian fitting for the bright pixels. The final 3-D data cube was constructed with an angular resolution parameter $N_{side}$ = 64, corresponding to a pixel size of $\sim$55.0 arcmin, and a spectral resolution of $\sim$0.92 \AA. Like the \textit{L}-band detector, the \textit{S}-band detector also suffered degradation over the period of the mission lifetime. We estimated the effective area as a function of time based on the median spectrum of the whole orbit data obtained with the sky survey mode. We used the whole orbit data instead of the standard calibration sources because the sensitivity of the \textit{S}-band, which was lower than that of the \textit{L}-band by an order of magnitude, was not good enough to use the survey observation of the calibration sources with limited exposure time for a calibration purpose. Figure \ref{fig1}a shows the variation of median photon count-rates over time. Out of 967 orbits, we selected and plotted 192 orbits, whose data have an average signal-to-noise ratio greater than 3.0. It can be seen that the \textit{S}-band detector sensitivity decreased drastically (by $\sim$44\%) during the first four months and gradually (by $\sim$9\%) over the next seven months; hence, it is reduced by $\sim$53\% over a year. We note that the detector sensitivity decreased by up to 30\% in the case of the \textit{L}-band over the observation period \citep{Jo2017}. We calculated the effective area for the entire observation period, which also varies according to the wavelength, using the two calibration sources HD 1337 and HD 93521 observed in December 2003 and April 2004, respectively, and through the interpolation based on the sensitivity degradation, as obtained from the whole orbit data. Examples of effective areas for several periods are shown in Figure \ref{fig1}b.

\begin{figure}[t]
	\begin{centering}
		\medskip{}
		\par\end{centering}
	\begin{centering}
		\includegraphics[clip,scale=0.6]{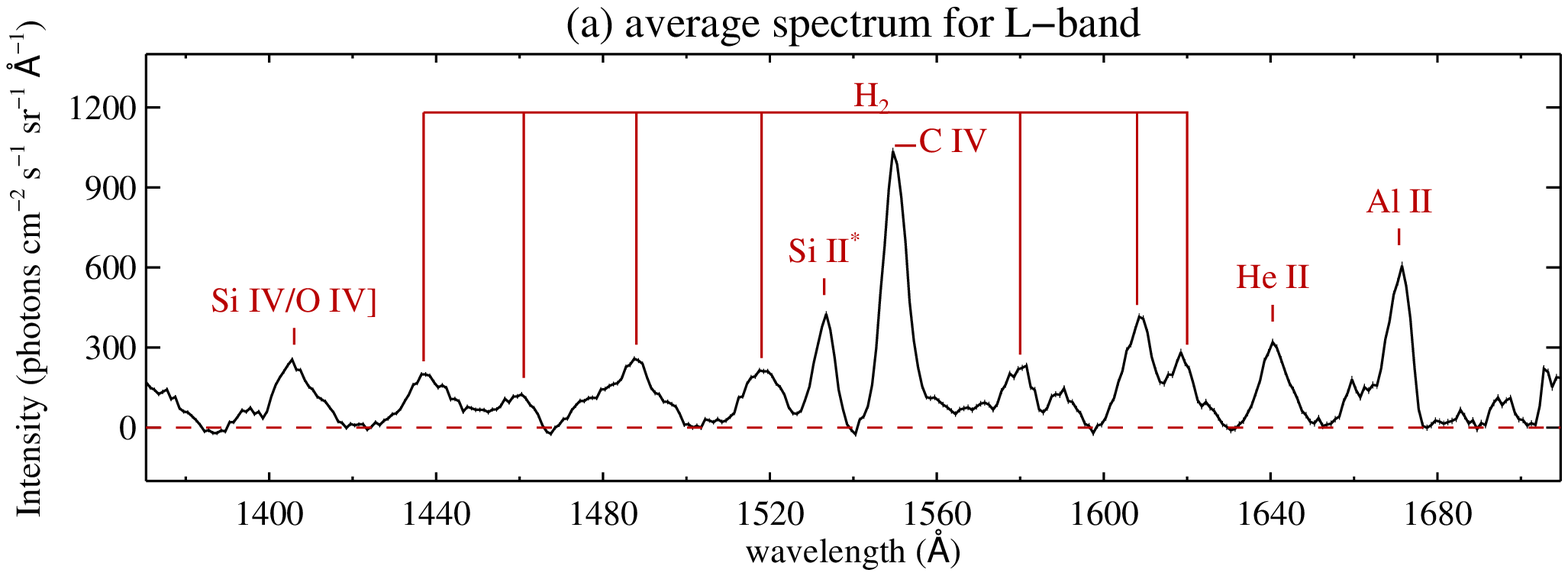}
		\par\end{centering}
	\begin{centering}
		\medskip{}
		\par\end{centering}
	\begin{centering}
		\includegraphics[clip,scale=0.6]{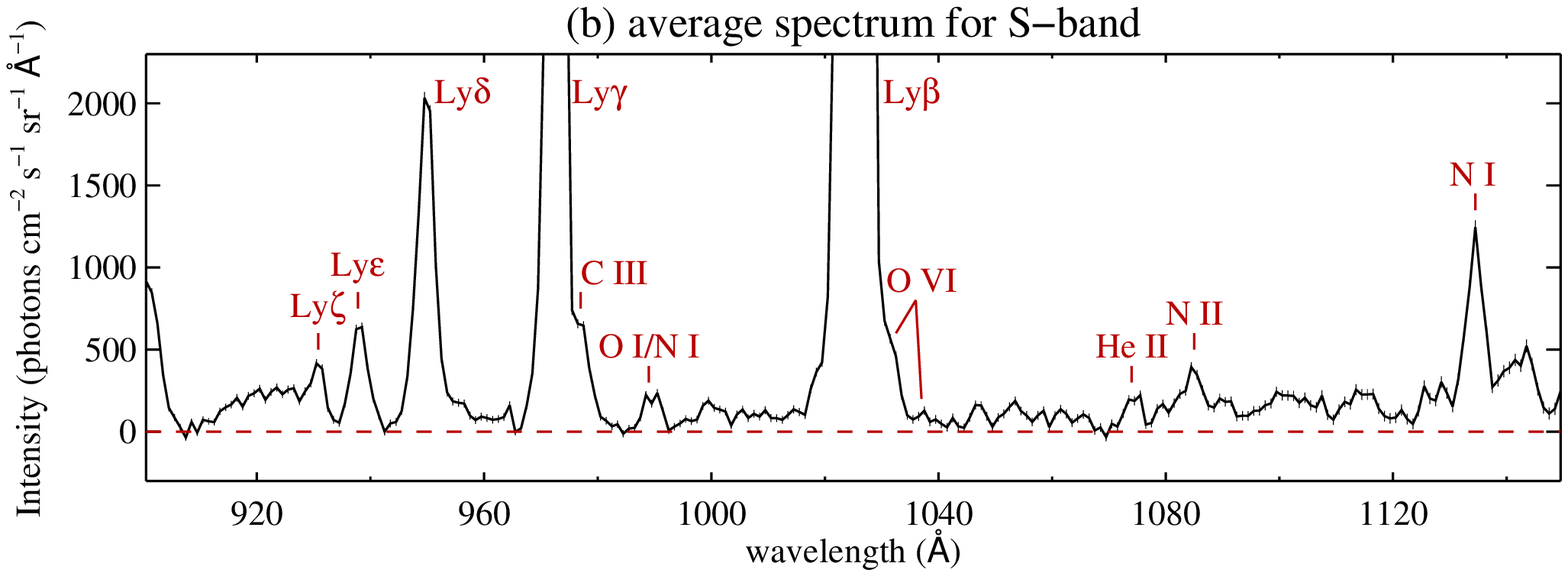}
		\par\end{centering}
	\begin{centering}
		\medskip{}
		\par\end{centering}
	\caption{\label{fig2}FIMS/SPEAR all-sky spectrum averaged over exposure time for the \textit{L}-band (upper) and \textit{S}-band (lower) channels: the Lyman series lines in the bottom panel are airglow lines.}
\end{figure}

Figure \ref{fig2} shows the exposure time-weighted and continuum-subtracted FUV spectrum for the \textit{L}-band and \textit{S}-band channels, whose intensity is given in the continuum unit (CU; photons cm$^{-1}$ sr$^{-1}$ \AA$^{-1}$ s$^{-1}$). A number of prominent emission lines are identified and marked in red, including interstellar atomic lines and $H_2$ fluorescence emission features, as well as the Lyman series airglow lines. Each of the interstellar emission lines may represent a particular state of ISM, for example, $H_2$ emission lines are observed primarily in molecular clouds with temperatures below $\sim$1,000 K, whereas \ion{Al}{2} and \ion{Si}{2}* emission lines are generated from warm neutral/ionized media with temperatures below $\sim$10,000 K. As we are interested in the cooling process of the interstellar hot gas, this study focuses on the \ion{C}{4} and \ion{O}{6} emission lines, which are the prominent indicators of ISM with temperatures of $\sim$$10^{5.0}$ K and $\sim$$10^{5.5}$ K, respectively. Furthermore, assuming that they originate from the same medium in thermal equilibrium, the intensity ratio of the two emission lines can be used to estimate the temperature of the transition-temperature gas. The \ion{Si}{4} emission lines ($\lambda\lambda$1394,1403 \AA) are also strong cooling lines, but they were excluded from this study owing to difficulties in separating them from the close \ion{O}{4} line ($\lambda$1400 \AA) in the present spectrum of medium resolution.

\subsection{Construction of the \ion{C}{4} and \ion{O}{6} Emission Line Maps}

\begin{figure}[t]
	\begin{centering}
		\medskip{}
		\par\end{centering}
	\begin{centering}
		\includegraphics[clip,scale=0.8]{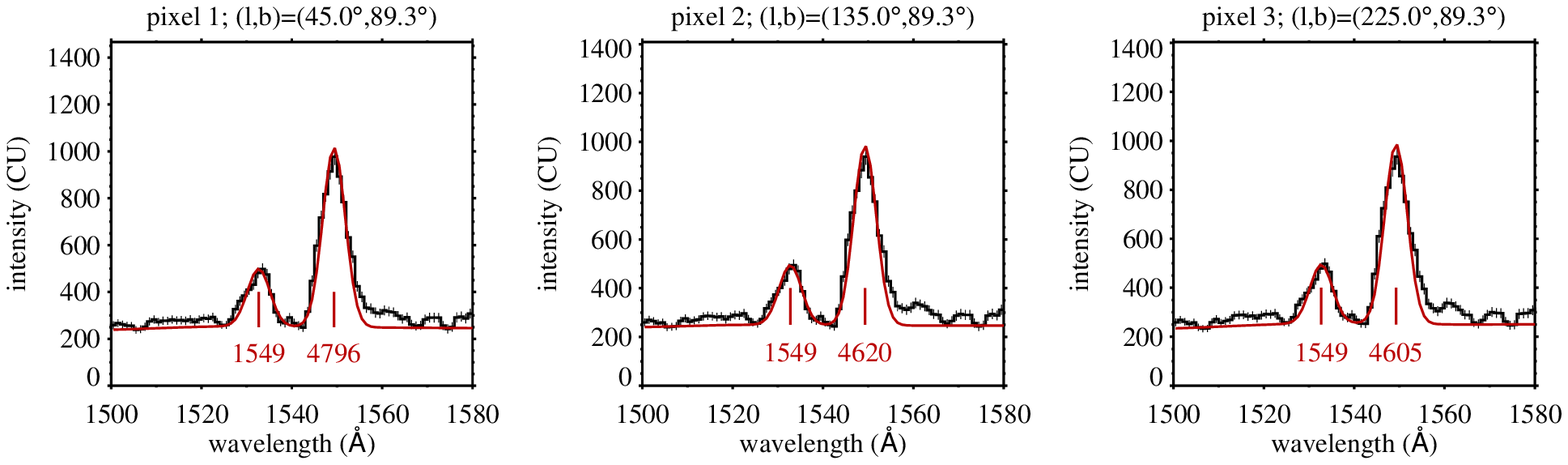}
		\par\end{centering}
	\begin{centering}
		\includegraphics[clip,scale=0.8]{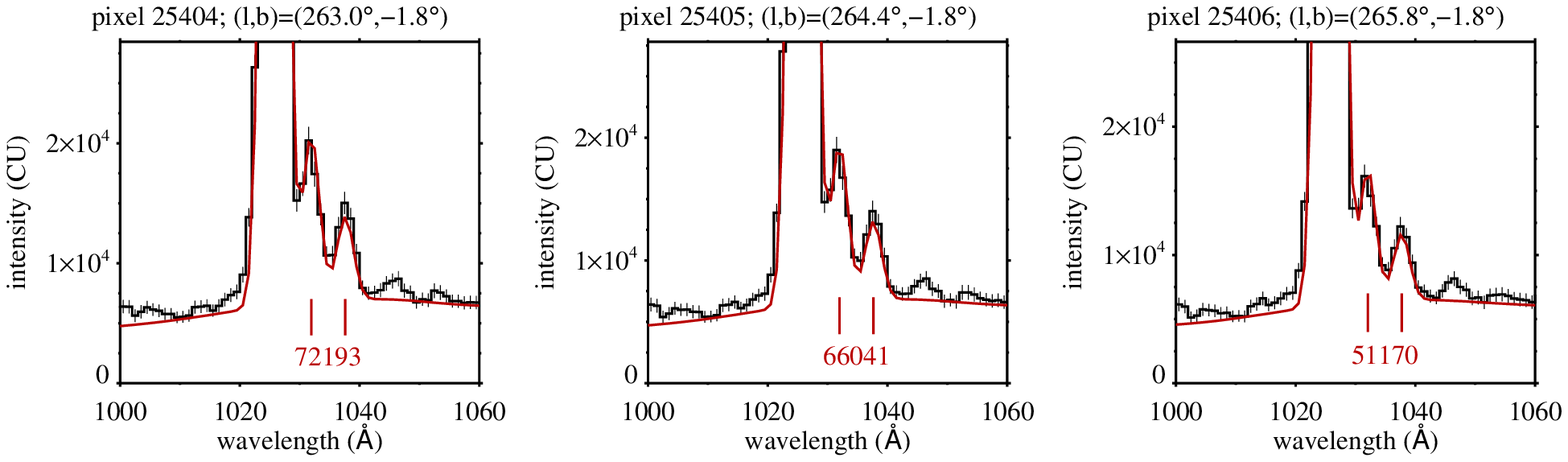}
		\par\end{centering}
	\begin{centering}
		\medskip{}
		\par\end{centering}
	\caption{\label{fig3}Example spectra with emission line fittings for the \textit{L}-band (upper panel) and the \textit{S}-band (lower panel). The black solid lines with error bars are the observed spectra, and the red solid lines are the fitted model spectra. The \textit{L}-band spectra include the \ion{Si}{2}* ($\lambda$1533 \AA) and \ion{C}{4} doublet ($\lambda\lambda$1548,1551 \AA) lines, and the \textit{S}-band spectra include the airglow Ly$\beta$ ($\lambda$1025 \AA) line and the \ion{O}{6} doublet ($\lambda\lambda$1032,1038 \AA) lines. The line intensities in he line unit (LU; photons cm$^{-2}$ sr$^{-1}$ s$^{-1}$) shown in red represent for the \ion{Si}{2}* and \ion{C}{4} lines in the upper panel of the \textit{L}-band, and the \ion{O}{6} line in the lower panel of the \textit{S}-band. The pixel number at the top of each plot indicates the identification number in the RING scheme of HEALPix for a resolution parameter of $N_{side}$ = 64 with the total number of pixels $N_{pixel}$=49152.}
\end{figure}

We have constructed \ion{C}{4} and \ion{O}{6} emission lines maps with a pixel size of $\sim$0.92\degr, corresponding to $N_{side} = 64$ in a HEALPix scheme \citep{Gor2005}, which is the same as those of the original data cubes.  However, we employed an adaptive smoothing method \citep[e.g.,][]{Seo2005} to keep the signal-to-noise ratio (SNR) above 7.0 for the wavelength bin of 1 \AA{}. A two-dimensional Gaussian function was used as a smoothing kernel with its full width at half maximum (FWHM) varying from 3\degr{} to 10\degr{} at 0.1\degr{} intervals. In this way, the angular size and the number of pixels are preserved, but the spatial resolution varies depending on the kernel size adopted at each pixel. We ignored the correlation between adjacent pixels caused by the adaptive smoothing because we are mainly interested in modeling the global distribution of the transition-temperature gas. Figure \ref{fig3} shows some examples of the fitting results. In the figures, the observed spectra and the model spectra are shown as black and red solid lines, respectively.

The three plots in the upper panel of Figure \ref{fig3} show the fitting results of the \textit{L}-band spectra for the three locations near the north Galactic pole. As can be seen from the figures, fitting was performed with two Gaussian functions centered around the peaks of the \ion{Si}{2}* and \ion{C}{4} emission features for the wavelength range 1500–1580 \AA. The \ion{C}{4} doublet ($\lambda\lambda$1548,1551 \AA) was fitted with a single Gaussian model because they are not readily resolved with the \textit{L}-band spectral resolution of $\sim$3 \AA. The weak emission features corresponding to the $H_2$ fluorescence were treated as a background continuum. The estimated line intensities of the \ion{Si}{2}* and \ion{C}{4} lines are given in line units (LU; photons cm$^{-2}$ sr$^{-1}$ s$^{-1}$) and shown in red just below each of these lines. The fitting for the \textit{S}-band spectra was performed in a similar manner for the wavelength range 1000–1060 \AA, where the \ion{O}{6} doublet ($\lambda\lambda$1032,1038 \AA) is located at the center. However, the region also contains a strong Ly$\beta$ airglow line ($\lambda$1025 \AA), which partially overlaps and generally dominates the \ion{O}{6} doublet lines. Hence, it is essential to have a sufficient SNR, which, in this study, was kept above 7.0 by employing adaptive smoothing with variable smoothing kernels, to identify the \ion{O}{6} doublet lines appropriately. As the two lines of the \ion{O}{6} doublet are distinguishable with the \textit{S}-band spectral resolution of $\sim$1.5 \AA, three Gaussian functions were employed to fit the spectra, including the one for the Ly$\beta$ airglow line. For the \ion{O}{6} doublet, the line ratio of I(1032)/I(1038) was set at 2, taking into account the collision-strength ratio. The three plots in the lower panel of Figure \ref{fig3} show the fitting results for the three locations around the Vela supernova remnant, which is one of the regions with bright \ion{O}{6} emissions. The total intensities of the \ion{O}{6} doublet are given in LU and shown in red just below the \ion{O}{6} doublet features.

\begin{figure}[t]
	\begin{centering}
		\medskip{}
		\par\end{centering}
	\begin{centering}
		\includegraphics[clip,scale=0.6,angle=90]{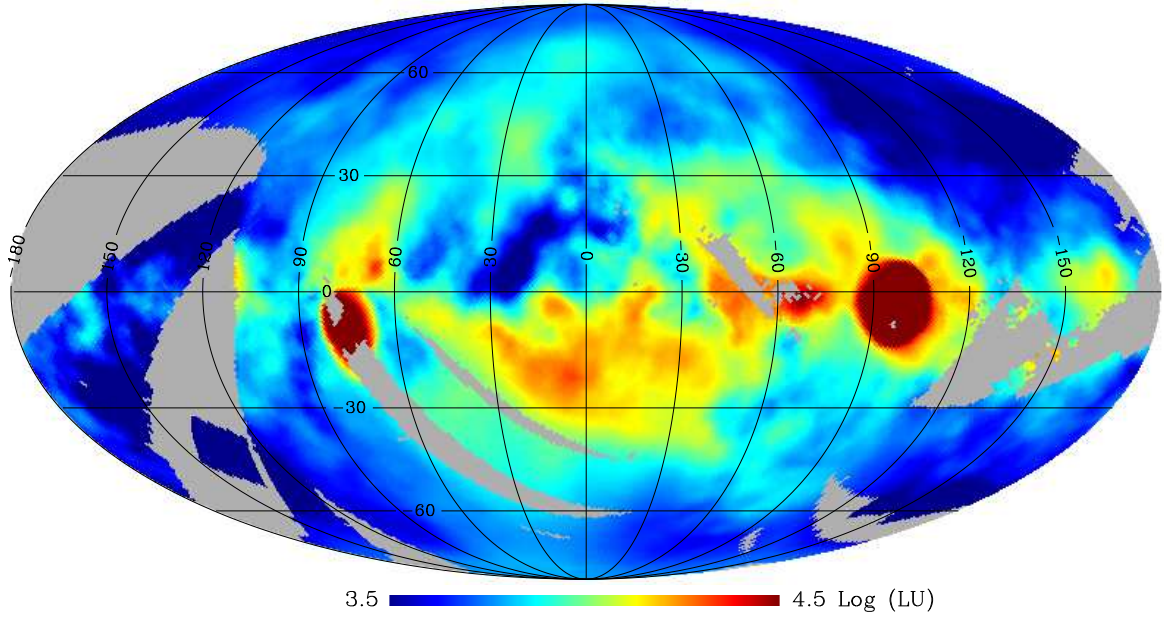}
		\par\end{centering}
	\begin{centering}
		\par\end{centering}
	\begin{centering}
		\includegraphics[clip,scale=0.6,angle=90]{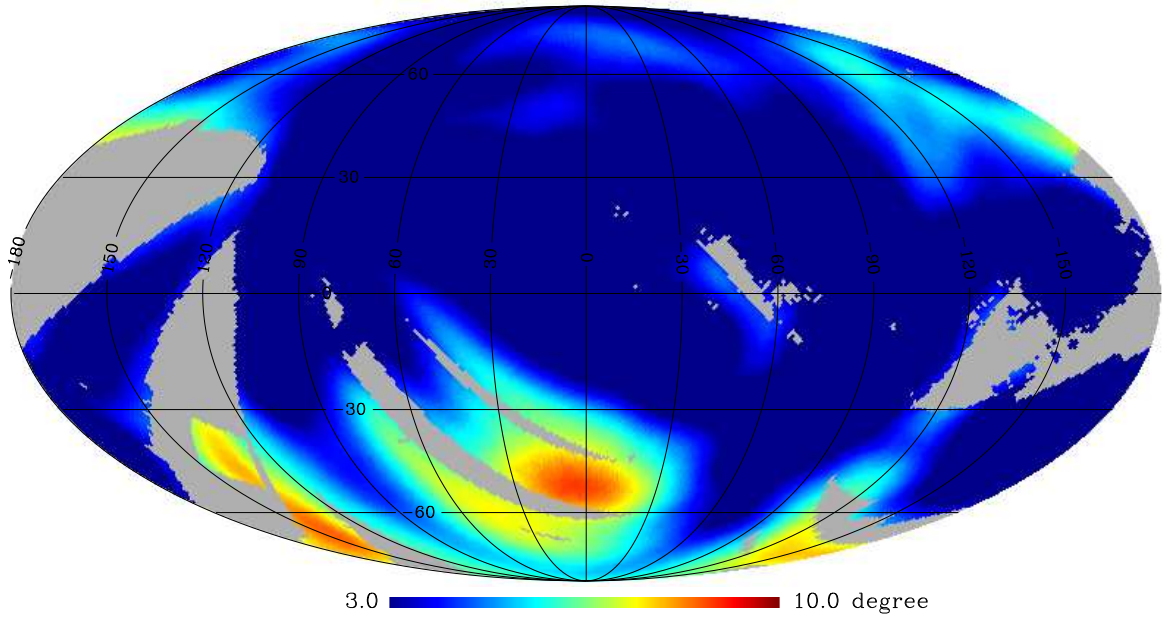}
		\par\end{centering}
	\begin{centering}
		\medskip{}
		\par\end{centering}
	\caption{\label{fig4}(a) \ion{C}{4} map on a logarithmic scale of LU, (b) FWHM map of the two-dimensional Gaussian kernel for the \ion{C}{4} map, (c) \ion{O}{6} map on a logarithmic scale of LU, and (d) FWHM map of the two-dimensional Gaussian kernel for the \ion{O}{6} map. The gray areas represent the regions where observations were not made or the SNR was low.}
\end{figure}

The resulting emission line maps are depicted in Figure \ref{fig4}. Figure \ref{fig4}a depicts the \ion{C}{4} intensity map and the corresponding smoothing radius (FWHM) of the 2D Gaussian kernel is shown in Figure \ref{fig4}b; Figure \ref{fig4}c shows the \ion{O}{6} intensity map and the corresponding smoothing radius (FWHM) of the 2D Gaussian kernel is shown in Figure \ref{fig4}d. The \ion{C}{4} and \ion{O}{6} maps account for $\sim$85\% and $\sim$77\% of the sky, respectively. The coverage of the \ion{O}{6} map is smaller than that of the \ion{C}{4} map; the FOV of the \textit{S}-band which includes \ion{O}{6} is smaller than that of the \textit{L}-band slit which observed \ion{C}{4}. Furthermore, more \textit{S}-band data were discarded than \textit{L}-band data because of their lower SNR. To enhance the SNR, a larger smoothing radius was generally employed for the \textit{S}-band than for the \textit{L}-band. The mean values and standard deviations of the FWHM of the Gaussian smoothing kernel are 3.8\degr{} $\pm$ 1.3\degr{} and 5.8\degr{} $\pm$ 2.2\degr{} for the \ion{C}{4} and \ion{O}{6} maps, respectively.

Although the emission maps in Figure \ref{fig4} are significantly smoothed to improve the SNR, several very bright and large-scale structures such as North Polar Spur, Vela, and Cygnus loop can still be distinguished around (-84$\degr$, 60\degr), (-100\degr, -3\degr), and (74\degr, -9\degr) in Galactic coordinates, respectively. These targets have been analyzed in detail in the papers of \citet{Par2007}, \citet{Nis2006}, and \citet{Seo2006}, respectively \citep[see also,][]{Kim2012, Kim2014}. The Vela supernova remnant is prominent in both the \ion{C}{4} and \ion{O}{6} maps, but the Cygnus loop is bright only in the \ion{C}{4} map because most of the corresponding region in the \ion{O}{6} map was excluded from the \textit{S}-band data cube owing to the low SNR. Meanwhile, other notable features such as the Monogem ring and Orion-Eridanus, can be identified around (-159\degr, 8\degr) and (169\degr, -40\degr) in Galactic coordinates, respectively, in the \ion{O}{6} map; they have also been analyzed in detail in the papers of \citet{Kim2007} and \citet{Kre2006}, respectively. In Figure \ref{fig4}c, excluding these bright individual objects, the \ion{O}{6} intensity generally tends to be higher in the northern Galactic region than in the southern Galactic region. \citet{Sav2003} also discovered that the \ion{O}{6} column density is systematically enhanced in the northern sky relative to the southern sky. Small targets, which are not properly resolved in the present maps with rather large smoothing kernels, have also been analyzed in previous papers \citep{Shi2007, Kim2010a, Kim2010b}. We note that the bright targets mentioned above are, in general, located close to the Sun. For example, the Vela and the Cygnus loop supernova remnants are at a distance of 250 pc and 440 pc, respectively. As we are interested in the global distribution of hot gas on the Galactic scale rather than individual targets, we first performed extinction correction before further discussing the environment of the transition-temperature gas as photons of FUV wavelengths suffer strong attenuation by the interstellar dust grains.

\subsection{Extinction Correction of the FUV Emission Maps}

For an accurate estimation of the dust extinction of FUV photons, it is obvious that detailed 3-D distributions of the interstellar dust, as well as the FUV sources, are required. However, as they are unavailable, we made a simple assumption that the FUV sources and interstellar dust are uniformly mixed. For this simple case, the extinction-corrected line intensity ($I_{corr}$) is related to the observed line intensity (\textit{I}) as follows:
\begin{equation}
	I_{corr}=I\times\frac{\tau}{1-e^{-\tau}},\label{eq1}
\end{equation}
where $\tau$ is the optical depth along the line-of-sight \citep[e.g.,][]{Nat1984}. The optical depths of \ion{C}{4} and \ion{O}{6} lines are given by:
\begin{equation}
	\tau_{C IV}=7.3\times E(B-V) \; at \, \lambda\sim1550 \AA,\label{eq2}
\end{equation}
\begin{equation}
	\tau_{O VI}=11.8\times E(B-V) \; at \, \lambda\sim1035 \AA,\label{eq3}
\end{equation}
based on the Milky Way extinction curve of \citet{Wei2001} for $R_V$ = 3.1. We downloaded the \textit{E(B-V)} map \citep{Sch1998} from the Legacy Archive for Microwave Background Data Analysis, and pixelated it using the HEALPix scheme with a resolution parameter of $N_{side}$ = 64, corresponding to the pixel size of the FUV line maps. Then, the \textit{E(B-V)} map was convolved with the two-dimensional Gaussian kernels\footnote{\href{http://lambda.gsfc.nasa.gov/product/foreground/f\_products.cfm}{http://lambda.gsfc.nasa.gov/product/foreground/f\_products.cfm}}, whose smoothing size was the same as those of the corresponding \ion{C}{4} and \ion{O}{6} maps. Finally, the smoothed \textit{E(B-V)} map was converted to optical depth maps, which were utilized for the extinction-correction of the \ion{C}{4} and \ion{O}{6} emissions using Eq. (\ref{eq1}). Uniform mixing of the FUV sources and interstellar dust might not be a bad assumption at least for the halo region, although it cannot be applied to the optically thick disk regions. 

The resulting extinction-corrected \ion{C}{4} and \ion{O}{6} emission line maps are shown in Figure \ref{fig5}, clearly displaying strong enhancement of the \ion{C}{4} and \ion{O}{6} emissions in the Galactic plane, whereas the corresponding intensities before the extinction correction were comparable to or even smaller than those of middle and high latitudes. Furthermore, it is seen that the intensity ratio of the two lines I(\ion{O}{6})/I(\ion{C}{4}) increased by a factor of $\sim$1.6 in the optically thick Galactic disk region of \textit{E(B-V)} = 1 after extinction correction, whereas it increased only slightly (by a factor of $\sim$1.1) in the optically thin, high latitude region of \textit{E(B-V)} = 0.05.  Since the assumption of a uniformly mixed ISM could be strongly violated in the complex, optically thick regions of the Galactic plane, we have excluded the Galactic disk region of |\textit{b}| < 15\degr{} from further discussions. Since we are mainly interested in the global structure of the transition-temperature gas in the halo region, this exclusion is not a problem. Although the low latitude region of |\textit{b}| < 15\degr{} is still shown in the all-sky maps, we have derived all physical parameters only from the region of |\textit{b}| > 15\degr{}.

\section{Results}

\subsection{Galactic Morphology and Physical Properties of the Transition-temperature Gas Traced by \ion{C}{4} and \ion{O}{6}}

\begin{figure}[t]
	\begin{centering}
		\medskip{}
		\par\end{centering}
	\begin{centering}
		\includegraphics[clip,scale=0.6,angle=90]{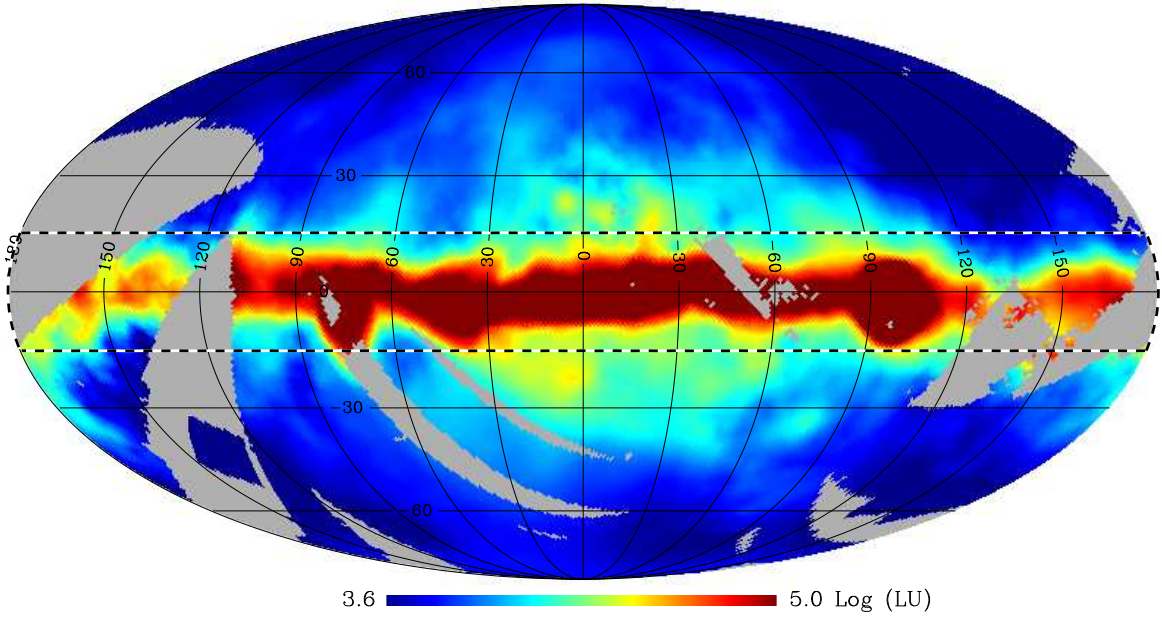}
		\par\end{centering}
	\begin{centering}
		\includegraphics[clip,scale=0.6,angle=90]{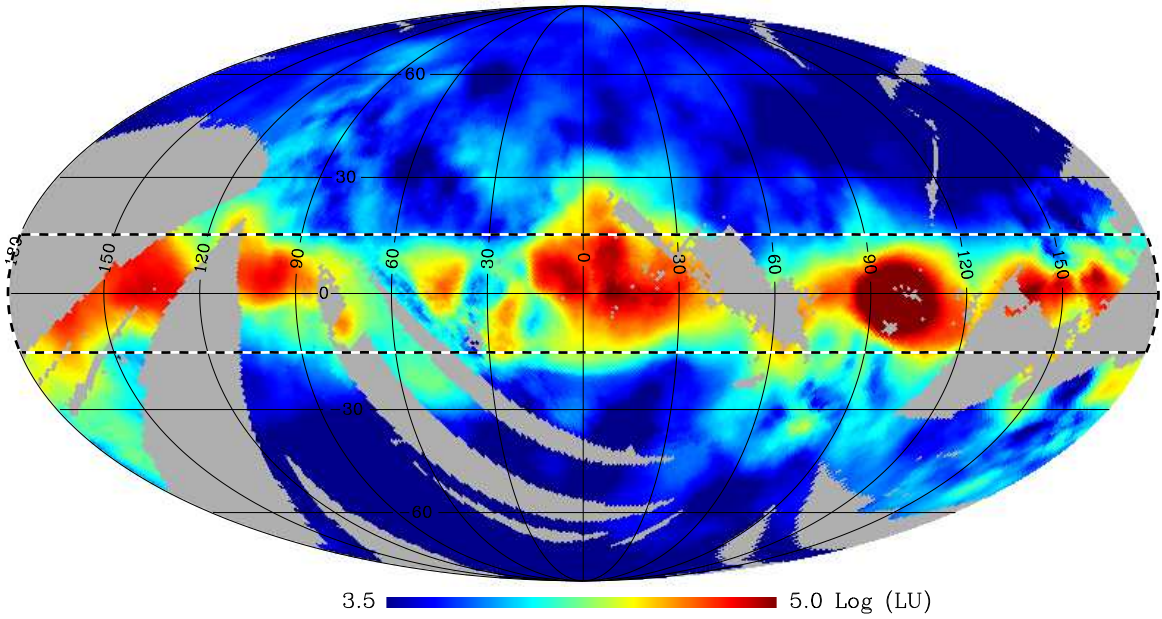}
		\par\end{centering}
	\begin{centering}
		\medskip{}
		\par\end{centering}
	\caption{\label{fig5}(a) \ion{C}{4} map and (b) \ion{O}{6} map after correcting for dust extinction. The strong extinction region of $|b|<15\degr$ around the Galactic disk is indicated with a dashed rectangular box.}
\end{figure}

\begin{figure*}[t]
	\begin{centering}
		\medskip{}
		\par\end{centering}
	\begin{centering}
		\includegraphics[clip,scale=0.7]{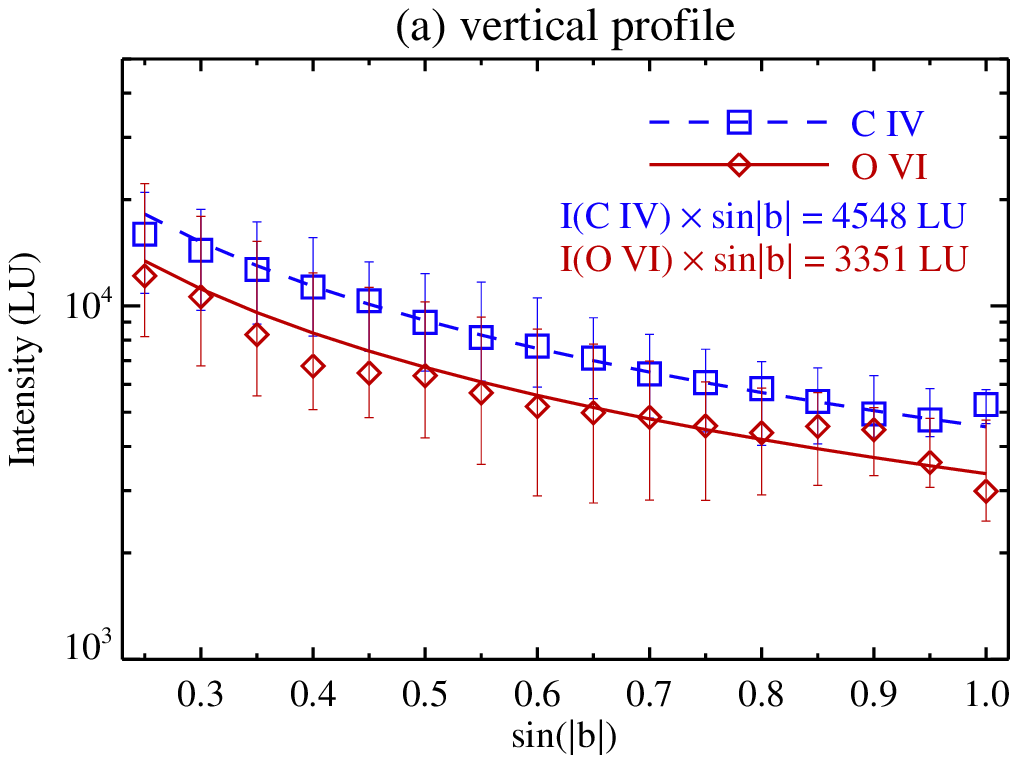}\
		\includegraphics[clip,scale=0.7]{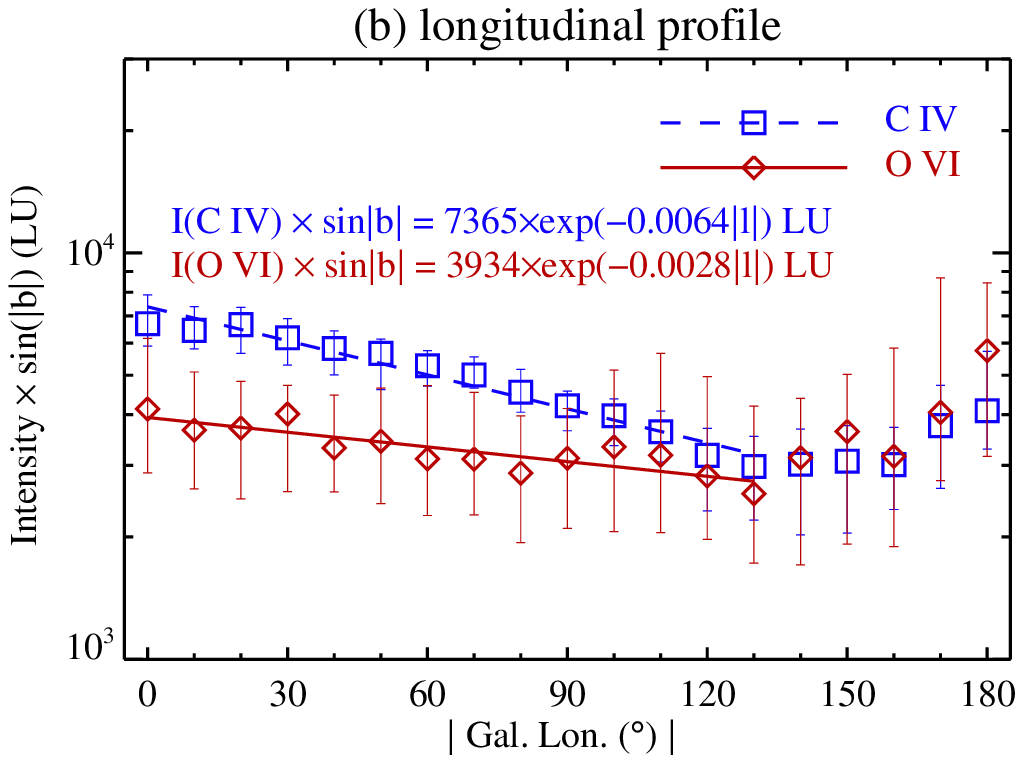}
		\par\end{centering}
	\begin{centering}
		\medskip{}
		\par\end{centering}
	\caption{\label{fig6}Intensity variation of \ion{C}{4} and \ion{O}{6} emissions; (a) Against Galactic latitudes, and (b) Against Galactic longitudes. The Galactic latitudes are binned with 0.05 steps of $\sin |b|$ in (a) and the longitudes are binned with 10\degr{} steps in (b). The error bars are  the upper and lower quartiles about the median values of the data of each bin. The functions shown in the panels represent the fitted lines of the data shown in the same panels.}
\end{figure*}

Using the global maps of Figure \ref{fig5}, we would like to discuss the variation of the \ion{C}{4} and \ion{O}{6} intensities on a Galactic scale. We plotted intensity variations, shown in Figure \ref{fig6}, with Galactic latitudes and longitudes by taking median values for each bin. Taking median values may reduce the influence of individual bright targets more effectively than taking averages. The error bars shown in Figure \ref{fig6} represent the upper and lower quartiles about the median values taken in the longitudinal (Figure \ref{fig6}a) or latitudinal (Figure \ref{fig6}b) direction, implying the spatial variation in the respective coordinate direction. The fitting errors of the line intensities for individual pixels were not considered in calculating error bars for the ensuing analysis because they are much smaller compared to the spatial variation of the intensities, as can be seen in the examples of the fitting results of Figure \ref{fig3}. Furthermore, we do not distinguish between the northern and the southern hemispheres. Though the \ion{O}{6} intensity is slightly higher in the northern hemisphere than in the southern hemisphere, the difference is much smaller than the errors associated with the fluctuations across the whole sky. 

Figure \ref{fig6}a shows the vertical intensity profiles plotted against $\sin{|b|}$, where \textit{b} is the Galactic latitude. The bin size of $\sin{|b|}$ is 0.05. The blue squares represent \ion{C}{4}, and the red diamonds represent \ion{O}{6}. As expected, it is seen that both the \ion{C}{4} and \ion{O}{6} intensities decrease toward the Galactic pole. \citet{Seo2011} showed that the FUV continuum emission, obtained from the same FIMS/SPEAR data, is well-described by a simple plane parallel model. We tested the same plane parallel model for the \ion{C}{4} and \ion{O}{6} intensities and plotted the best-fit models, as in Figure \ref{fig6}a. The blue dashed line and the red solid line correspond to the \ion{C}{4} and \ion{O}{6} intensities, respectively. It is seen that the two FUV emission lines are described well by a simple plane parallel model. The values of $I\times\sin{|b|}$, equal to the emissivity multiplied by a scale height if an exponential model is assumed, are approximately 4500 LU and 3400 LU for the \ion{C}{4} and \ion{O}{6}, respectively.

We also plotted the longitudinal intensity profiles against the Galactic longitudes, shown in Figure \ref{fig6}b. The symbols and colors are the same as those in Figure \ref{fig6}a. To eliminate the cosecant effect associated with Galactic latitudes, the median value of $I\times\sin{|b|}$ is shown in Figure \ref{fig6}b for each Galactic longitude bin of size 10\degr. It is clearly seen that both the \ion{C}{4} and \ion{O}{6} intensities decrease with an increase in |\textit{l}|, except for the region of $|l| > 130\degr$, which includes the bright local object Orion-Eridanus that extends to $b\sim -50\degr$ and thus may have affected the result. Excluding the region of $|l| > 130\degr$, we fitted the intensity profiles of \ion{C}{4} and \ion{O}{6} with exponential functions, as represented by the blue dashed line and the red solid line in Figure \ref{fig6}b, respectively. We can see that both the \ion{C}{4} and \ion{O}{6} intensities decrease with increasing |\textit{l}|, implying that the distribution of the transition-temperature gas depends on the Galacto-centric radius. Furthermore, the \ion{C}{4} intensity decreases faster than the \ion{O}{6} intensity. The \ion{C}{4} intensity reduced by $\sim$57\% at $l=130\degr$ from its value at $l=0\degr$, whereas the \ion{O}{6} intensity reduced by $\sim$31\%. This may imply that the gas temperature and/or gas density depends on the Galacto-centric radius.

ISM is never in exact thermodynamic equilibrium nor in a strict steady state but, instead, it constantly evolves by interacting with its environment. Nevertheless, a simple model based on static thermal equilibrium may still provide illuminating information on the physical properties of ISM. Assuming optically thin and collisionally excited plasma, the intensity of the emission line from the ion species \textit{i} is given by:
\begin{equation}
	I_{i}=\frac{1}{4\pi}\int_{0}^{\infty} n_i n_e \langle\sigma v\rangle_e d{x},\label{eq4}
\end{equation}
where $n_i$ and $n_e$ are the ion and electron densities, and $\langle\sigma v\rangle_e$ is the electron-impact excitation rate coefficient corresponding to the emission line in consideration. The integration was performed along the line of sight. For simplicity, we further assumed that the ISM is isothermal plasma with a constant temperature along the line of sight. Then, the above equation can be reduced to:
\begin{equation}
	I_{i}=\frac{\langle\sigma v\rangle_e}{4\pi}\frac{n_i}{n_H}\frac{n_H}{n_e}\int_{0}^{\infty} n_e^2 d{x}=\gamma_i \frac{n_i}{n_H}\frac{n_H}{n_e} EM_i, \label{eq5}
\end{equation}
with the following definition of the emission measure $EM_i$:
\begin{equation}
	EM_{i}=\int_{0}^{\infty} n_e^2 d{x}, \label{eq6}
\end{equation}
As $EM_i$ is the same for all ion emission lines if they originate from the same plasma, we will use EM instead of $EM_i$ without the subscript indicating a specific ion emission line. Strictly speaking, the \ion{O}{6} and \ion{C}{4} emissions may not originate from the same gas parcel. However, this assumption would not be too unrealistic if highly ionized atoms such as \ion{O}{6} and \ion{C}{4} are mainly located in adjacent regions where the hot gas is being cooled. In addition, \citet{Sav2009} showed that highly ionized atoms, such as \ion{O}{6}, \ion{N}{5}, \ion{C}{4}, \ion{Si}{4}, and \ion{Fe}{3}, are distributed with a similar scale height of approximately 3 kpc. \citet{Wak2012} suggested that the most likely explanation for this similarity is that the \ion{O}{6} is associated with the same structures from which \ion{C}{4} and \ion{Si}{4} originate. We therefore assume that the \ion{O}{6} and \ion{C}{4} emissions originate from the same gas parcel. The first three terms on the right-hand side of Eq. (\ref{eq5}) are functions of the ion abundance and gas temperature. Hence, by observing the ion line intensity $I_i$, we can determine the gas temperature with a suitable choice of ion abundance. For example, the ion abundance $n_i/n_H$ can be taken to be the ionization fraction at the solar abundance given in \citet{Sut1993}, in which the collisional ionization equilibrium (CIE) was assumed, and the electron density $n_e$ can be set as 1.2$n_H$ for a fully ionized gas. Furthermore, the $\gamma_i$, as defined by $\langle\sigma v\rangle_e / 4\pi$ in Eq. (\ref{eq5}), can be expressed according to \citet{Shu1994} as follows:
\begin{equation}
	\gamma_{C IV}=1.09\times 10^{-8} \cdot T_5^{-0.5} \cdot e^{\frac{-0.929}{T_5}} \cdot \left[\frac{\Omega_{C IV}}{10}\right], \label{eq7}
\end{equation}
\begin{equation}
	\gamma_{O VI}=5.43\times 10^{-9} \cdot T_5^{-0.5} \cdot e^{\frac{-1.392}{T_5}} \cdot \left[\frac{\Omega_{O VI}}{5}\right], \label{eq8}
\end{equation}
where $T_5$ is the gas temperature in units of $10^5$ K. The Maxwellian-averaged collision strength of the line emissions \ion{C}{4} and \ion{O}{6} are, respectively $\Omega_{C IV}$ and $\Omega_{O VI}$, which depend on the gas temperature. We adopted these values from \citet{Shu1994} and \citet{Tay2003}.

\begin{figure}[t]
	\begin{centering}
		\medskip{}
		\par\end{centering}
	\begin{centering}
		\includegraphics[clip,scale=0.7]{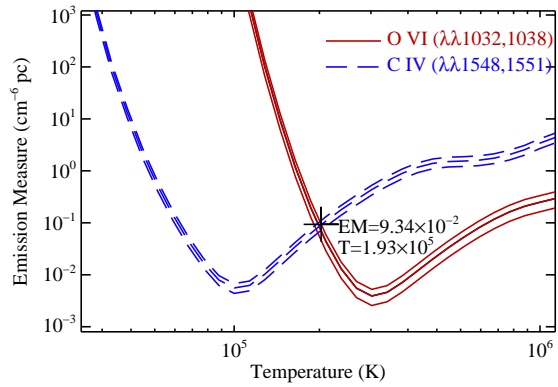}
		\par\end{centering}
	\begin{centering}
		\medskip{}
		\par\end{centering}
	\caption{\label{fig7}Example for the estimation of gas temperature and emission measure from the observed intensities of \ion{C}{4} and \ion{O}{6} lines. The curves shown here are obtained for the Galactic pole region of $b>60\degr$. The three curves for each \ion{C}{4} and \ion{O}{6} correspond to the maximum, average, and the minimum intensities within the 1-$\sigma$ confidence range, from the top.}
\end{figure}

Figure \ref{fig7} shows an example of the curves in the parameter space of the gas temperature and emission measure, constrained by the \ion{C}{4} and \ion{O}{6} intensities observed in the Galactic pole region of $b > 60\degr$. The intensities are 5206 $\pm$ 1109 LU for \ion{C}{4} and 4127 $\pm$ 1416 LU for \ion{O}{6}. The blue dashed lines and the red solid lines correspond to \ion{C}{4} and \ion{O}{6}, respectively. Three curves were plotted for each of the \ion{C}{4} and \ion{O}{6} lines, as shown in the figure. They correspond to the maximum, average, and the minimum values within the 1-sigma confidence range of the observed \ion{C}{4} and \ion{O}{6} intensities of the Galactic pole region, from the top. We note that the three curves for each of the two species are rather close to each other, implying that spatial variations of the density and temperature may not be significant, although the intensity variations of \ion{C}{4} and \ion{O}{6} are not that small. As both the \ion{C}{4} and \ion{O}{6} lines are assumed to originate from the same plasma, the temperature and EM are determined from the intersection of the two curves of \ion{C}{4} and \ion{O}{6}. They are $T = (1.93 \pm 0.07) \times 10^5$ K and $EM = (9.3 \pm 2.6) \times 10^{-2}$ cm$^{-6}$ pc, respectively.

We have constructed the gas temperature map and the emission measure map following the procedure described below. First, it can be shown from Eqs. (\ref{eq5})-(\ref{eq8}) that the intensity ratio of \ion{O}{6} to \ion{C}{4} is related to the gas temperature as follows:

\begin{equation}
	\frac{I_{O VI}}{I_{C IV}} = \frac{\gamma_{O VI}}{\gamma_{C IV}} \frac{n_{O VI}}{n_{C IV}} = \frac{\Omega_{O VI}}{\Omega_{C IV}} \frac{n_{O VI}}{n_{C IV}} \cdot e^{-\frac{46,288 K}{T}}, \label{eq9}
\end{equation}

\begin{figure}[t]
	\begin{centering}
		\medskip{}
		\par\end{centering}
	\begin{centering}
		\includegraphics[clip,scale=0.6,angle=90]{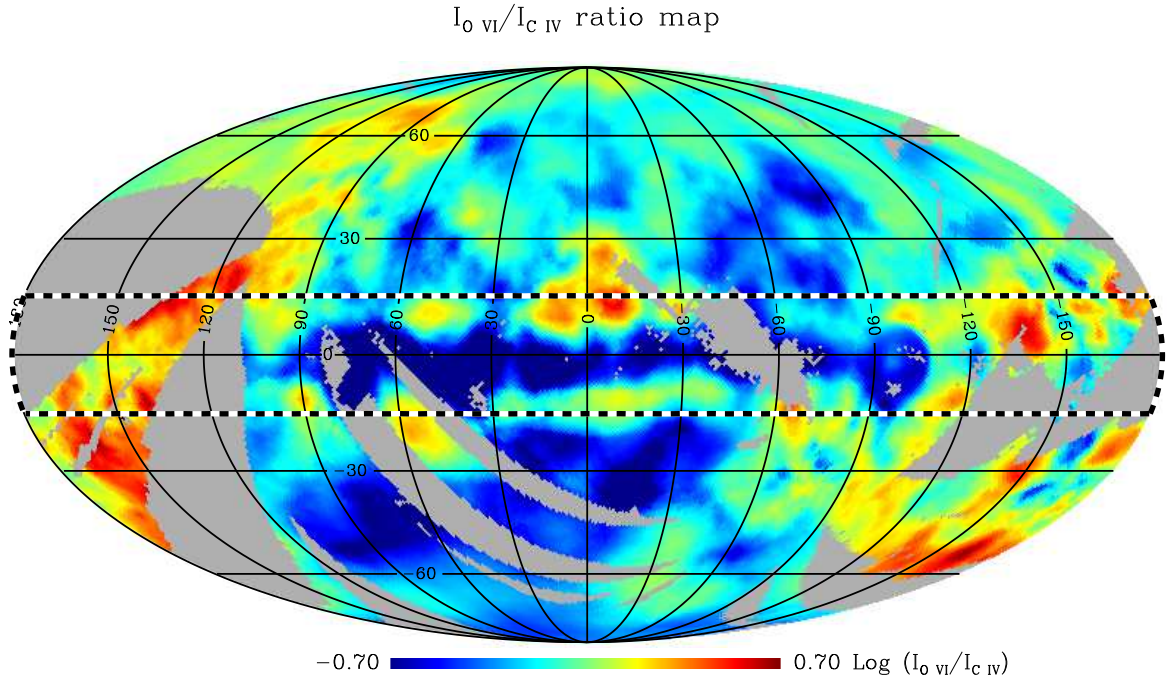}
		\par\end{centering}
	\begin{centering}
		\medskip{}
		\par\end{centering}
	\caption{\label{fig8}Intensity ratio map of $I_{O VI}/I_{C IV}$ .}
\end{figure}

\begin{figure}[t]
	\begin{centering}
		\medskip{}
		\par\end{centering}
	\begin{centering}
		\includegraphics[clip,scale=0.6,angle=90]{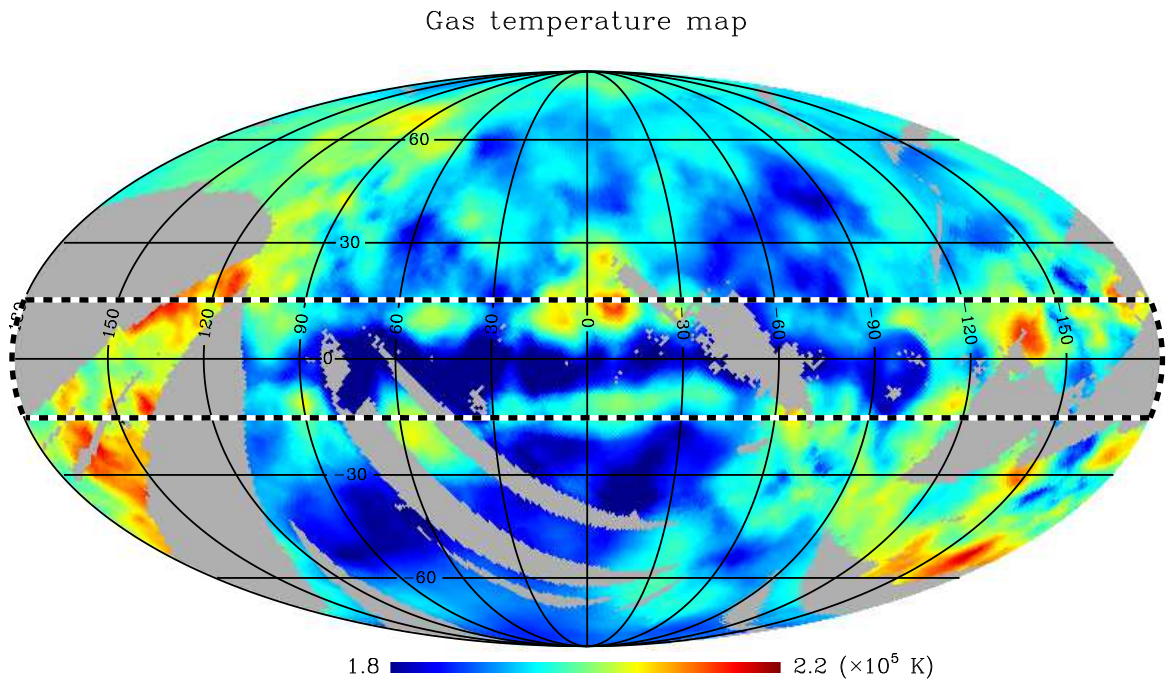}
		\par\end{centering}
	\begin{centering}
		\medskip{}
		\par\end{centering}
	\caption{\label{fig9}Gas temperature map derived from the intensity ratio $I_{O VI}/I_{C IV}$.}
\end{figure}

\begin{figure}[t]
	\begin{centering}
		\medskip{}
		\par\end{centering}
	\begin{centering}
		\includegraphics[clip,scale=0.9]{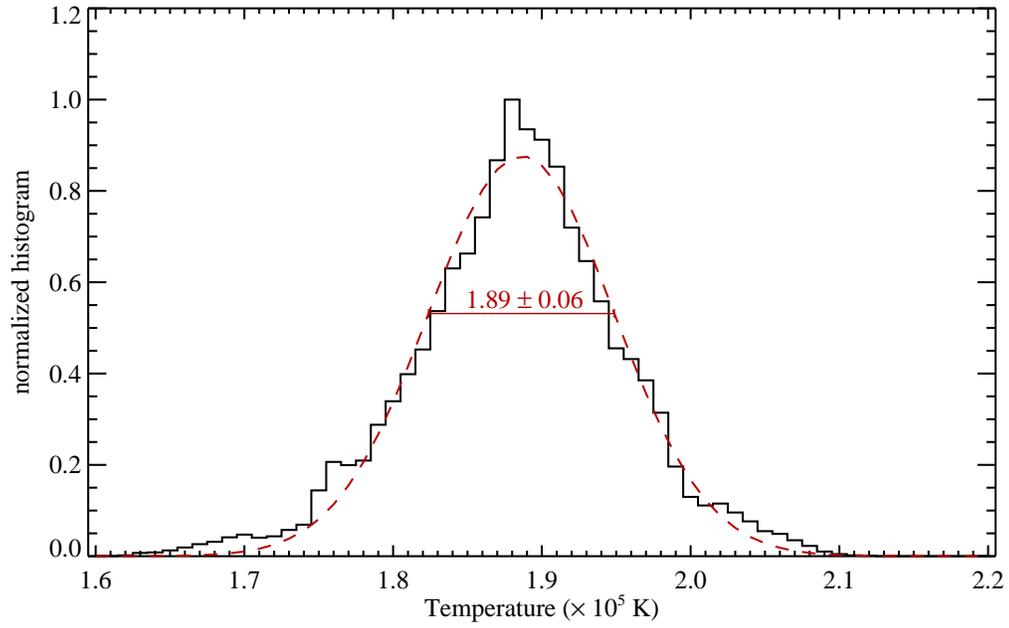}
		\par\end{centering}
	\begin{centering}
		\medskip{}
		\par\end{centering}
	\caption{\label{fig10} Normalized histogram of the estimated temperature for the pixels with |\textit{b}| > 15\degr{}. The red dashed line represents the best-fit Gaussian function.}
\end{figure}

\begin{figure}[t]
	\begin{centering}
		\medskip{}
		\par\end{centering}
	\begin{centering}
		\includegraphics[clip,scale=0.6,angle=90]{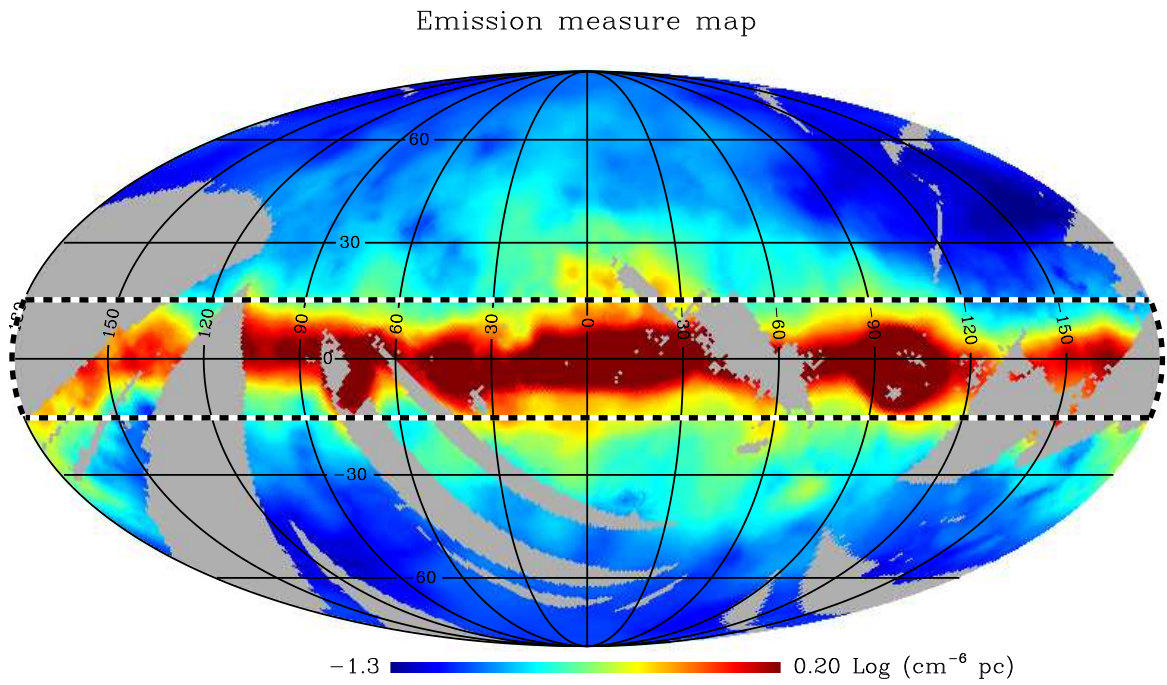}
		\par\end{centering}
	\begin{centering}
		\medskip{}
		\par\end{centering}
	\caption{\label{fig11}Emission measure map}
\end{figure}

The intensity ratio $I_{O VI}/I_{C IV}$, as shown in Figure \ref{fig8}, varies significantly from less than 0.04 up to 5, although its median value is 0.75 and it is less than 2.0 for more than 94\% of the observed sky. Considering the temperature dependence of the ion abundance under the assumption of CIE, Eq. (\ref{eq9}) determines the gas temperature uniquely from the intensity ratio of $I_{O VI}/I_{C IV}$ if it is smaller than 10. The resulting gas temperature map is shown in Figure \ref{fig9}. Despite the wide variation of $I_{O VI}/I_{C IV}$, the gas temperature is more or less uniform across the whole sky, with its value of (1.89 $\pm$ 0.06) $\times$ 10$^5$ K for the region of |\textit{b}| > 15\degr{}, in which the error bar is the standard deviation about the mean. Depicted in Figure \ref{fig10} is the normalized histogram of the estimated temperature for individual pixels. We note that the histogram closely matches the Gaussian distribution with a narrow FWHM (full width at half maximum) value of 0.12 $\times$ 10$^5$ K, indicating the temperature is well defined within a narrow range. As can be inferred from Eq. (\ref{eq9}), the intensity ratio is very sensitive to changes in temperature, allowing only small range of temperatures to fit the observed \ion{C}{4} and \ion{O}{6} intensities. The spectral fitting errors, associated with the estimation of \ion{C}{4} and \ion{O}{6} intensities, were not taken into account because these errors are negligible compared to the spatial variation of the intensities. The EM can be derived by substituting either the \ion{O}{6} or the \ion{C}{4} intensity and the estimated gas temperature into Eq. (\ref{eq5}). The uncertainties of the \ion{C}{4} and \ion{O}{6} intensities in each pixel are negligible compared to their variation over the sky; the spatial variation of the intensities is predominant. Thus, we ignored their uncertainties in obtaining the EM map. The resulting map is shown in Figure \ref{fig11}, which resembles the intensity maps of \ion{C}{4} and \ion{O}{6} in Figure \ref{fig5} due to an approximate uniform gas temperature throughout the entire Galaxy.

\begin{figure*}[t]
	\begin{centering}
		\medskip{}
		\par\end{centering}
	\begin{centering}
		\includegraphics[clip,scale=0.7]{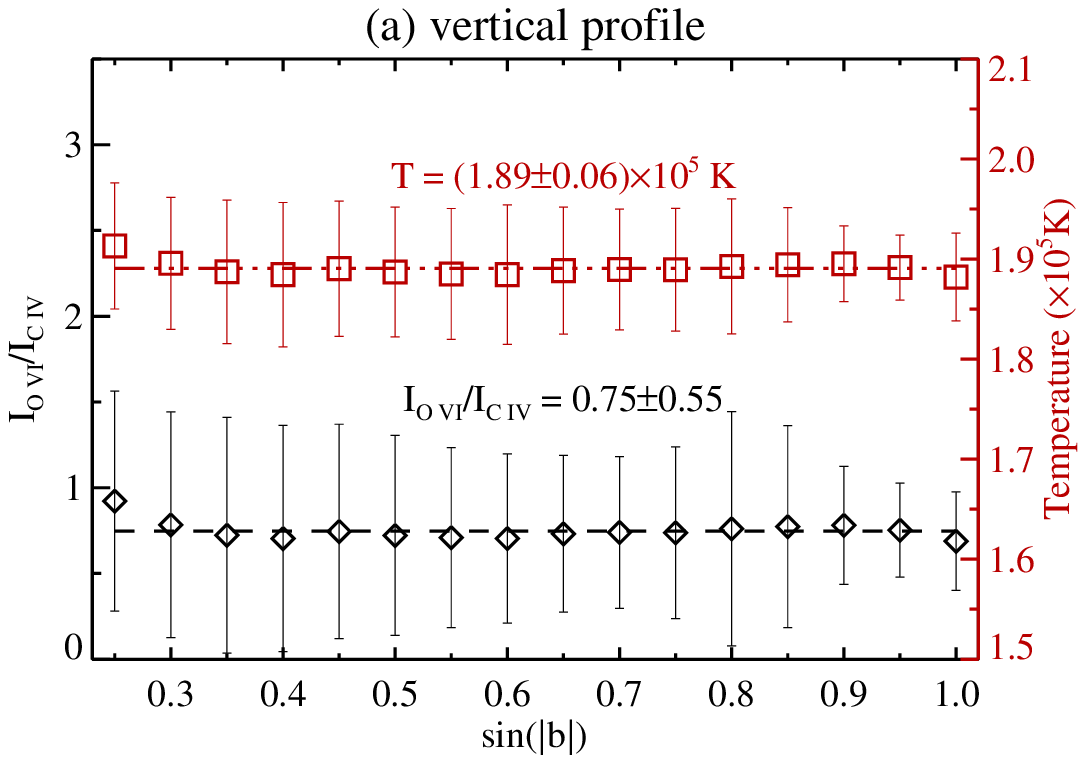}\
		\includegraphics[clip,scale=0.7]{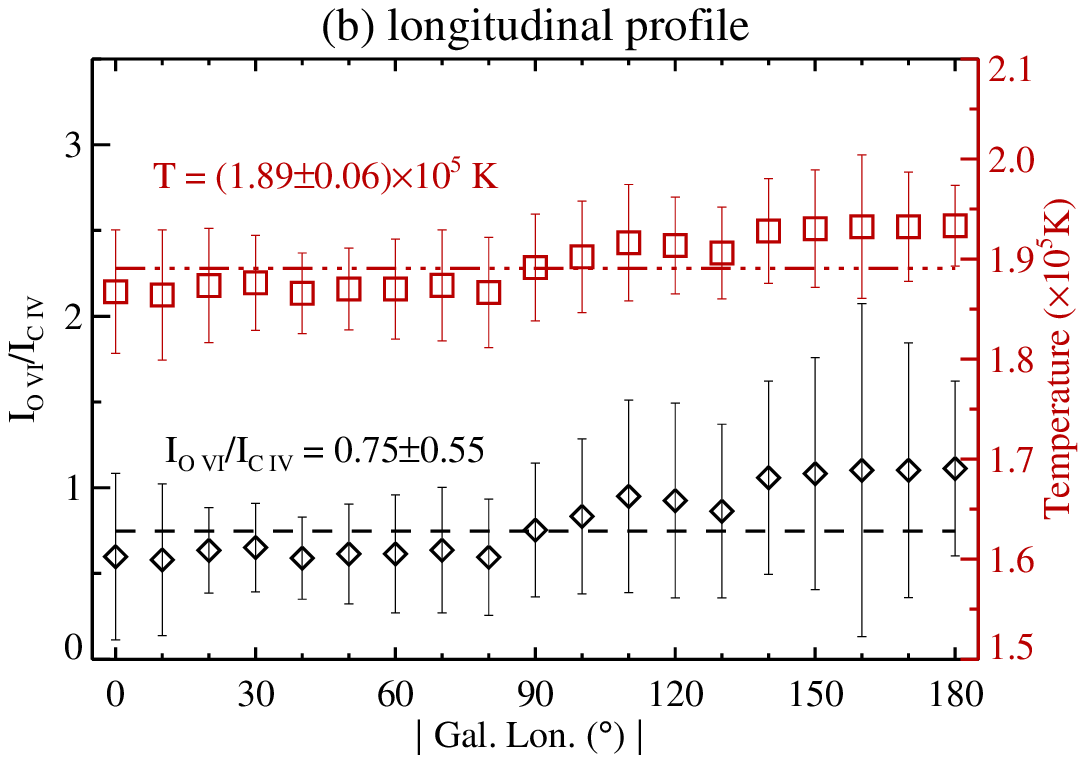}
		\par\end{centering}
	\begin{centering}
		\medskip{}
		\par\end{centering}
	\caption{\label{fig12}Variation of the intensity ratio $I_{O VI}/I_{C IV}$ and the gas temperature: (a) Along the Galactic latitude, and (b) Along the Galactic longitude. The error bars are the standard deviation of the data of each bin.}
\end{figure*}

To investigate the global variations of the intensity ratio $I_{O VI}/I_{C IV}$ and the gas temperature on the Galactic scale, we averaged them over the Galactic longitudes and latitudes to produce their vertical and longitudinal profiles, respectively, which may suppress the local variations seen in the all-sky maps. Again, we do not distinguish between the northern and the southern hemispheres, as in the case of the \ion{C}{4} and the \ion{O}{6} intensities shown in Figure \ref{fig6}, because the effects of the systematic difference between the two hemispheres are much smaller than the errors originating from the spatial fluctuations. The large-scale structures, such as North polar spur and Orion-Eridanus shown in Figure \ref{fig4}, may contribute to the local variations. We have chosen sufficiently large bins of Galactic coordinates, each comparable to the size of the individual large-scale structures, to minimize their influence. Hence, the exclusion or inclusion of these features does not alter the following analysis results significantly. Figure \ref{fig12}a shows the vertical profiles of the intensity ratio $I_{O VI}/I_{C IV}$ and the gas temperature plotted against $\sin{|b|}$. The black diamonds represent the intensity ratio, and the red squares represent the gas temperature, together with their standard deviations. The median values of the intensity ratio and the gas temperature obtained from all the pixels at |\textit{b}| > 15\degr{} are shown as horizontal lines. It is seen that both the intensity ratio and the gas temperature are almost uniform regardless of the Galactic latitude, even though the standard deviations are rather large, particularly for the case of the intensity ratio. Similarly, Figure \ref{fig12}b shows the longitudinal profiles of the intensity ratio and the gas temperature plotted against the Galactic longitude. The symbols and colors are the same as those in Figure \ref{fig12}a. It seems that the line ratio is larger and, as a consequence, the gas temperature is higher in the region of |\textit{l}| > 90\degr{} than in the region of |\textit{l}| < 90\degr{}; even though the difference is not very significant, and the error bars are rather large. Figure \ref{fig9} of the Galactic temperature distribution also generally confirms this radial dependence, though there exist local variations. Hence, we may suspect that the hot gases originating from the supernova explosions in the disk escape more easily toward the halo in the outer Galaxy region, producing higher temperature gas there than in the inner Galaxy region. In fact, the gas density is generally lower in the region of the outer Galaxy of |\textit{l}| > 90\degr{} than in the region of the inner Galaxy of |\textit{l}| < 90\degr{}, as will be discussed later in association with Figure \ref{fig14}.

\begin{figure*}[t]
	\begin{centering}
		\medskip{}
		\par\end{centering}
	\begin{centering}
		\includegraphics[clip,scale=0.7]{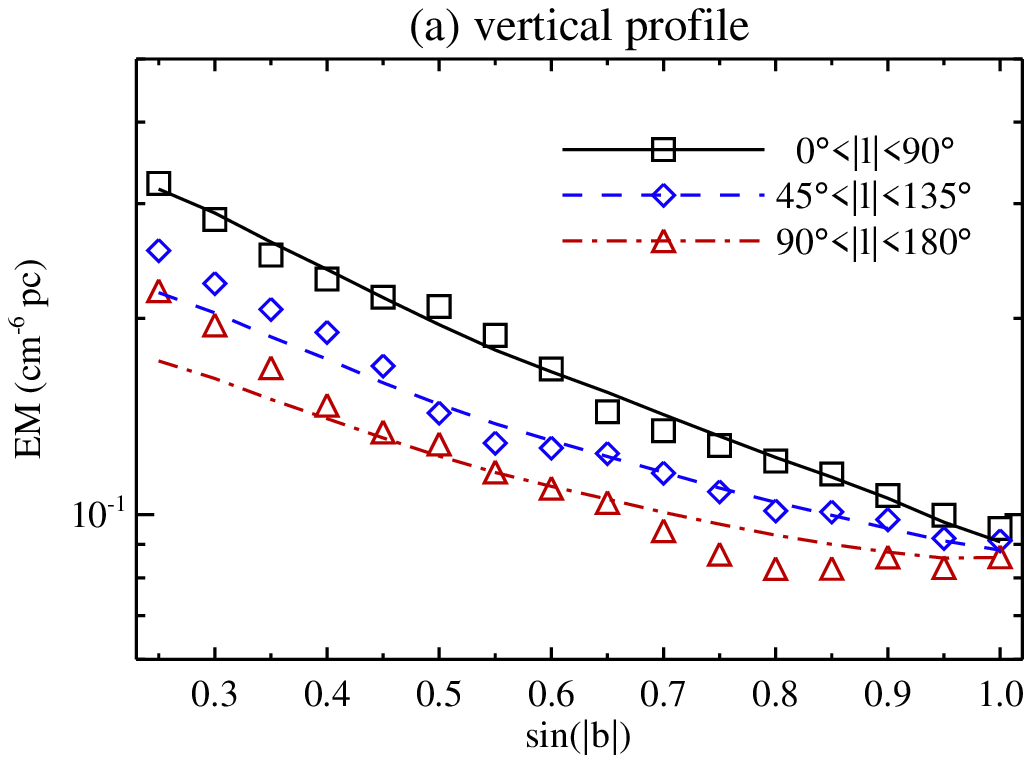}\
		\includegraphics[clip,scale=0.7]{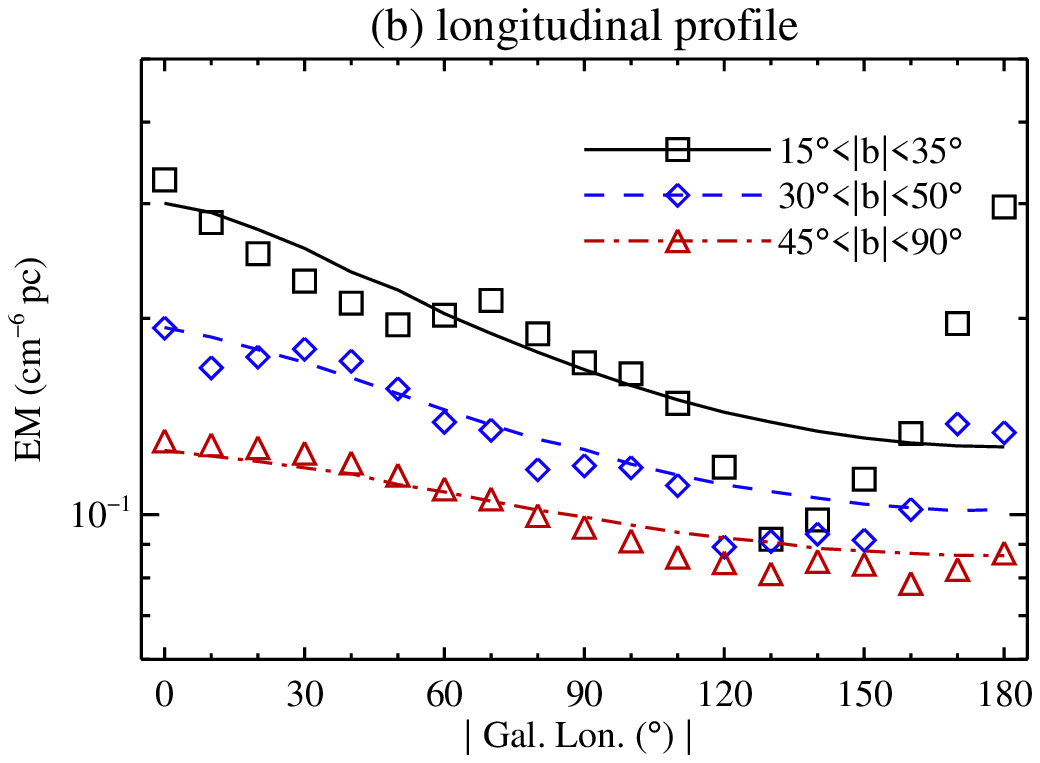}
		\par\end{centering}
	\begin{centering}
		\medskip{}
		\par\end{centering}
	\caption{\label{fig13}Variation of the emission measure (a) along the Galactic latitude and (b) along the Galactic longitude. The vertical profiles in (a) were obtained for three different longitudinal regions. The longitudinal profiles in (b) were obtained for three different latitudinal regions.}
\end{figure*}

Similarly, Figure \ref{fig13} presents the vertical and longitudinal profiles of the EM. However, since the EM varies significantly with changes in longitude and latitude, we divided the Galactic space into three subregions for each vertical and longitudinal profile. Figure \ref{fig13}a shows the vertical profiles for the ranges of 0\degr{} < |\textit{l}| < 90\degr{} (black squares), 45\degr{} < |\textit{l}| < 135\degr{} (blue diamonds), and 90\degr{} < |\textit{l}| < 180\degr{} (red triangles); and Figure \ref{fig13}b shows the longitudinal profiles for the ranges of 15\degr{} < |\textit{b}| < 35\degr{} (black squares), 30\degr{} < |\textit{b}| < 50\degr{} (blue diamonds), and 45\degr{} < |\textit{b}| < 90\degr{} (red triangles). We would like to fit the observed profiles of the EM with a model distribution of hot gas. Recently, there have been several studies on this topic based on the X-ray and ultraviolet observations. Notably, \citet{Yao2009} proposed a disk-like model and \citet{Mil2013, Mil2015} proposed a spherical model, while \citet{Nak2018} suggested a combination of the two. Although we tried both the spherical and disk-like models, we will describe the results of the disk-like model only because the spherical model resulted in extremely large $\chi^2$ values for all the parameters we explored.

For the disk-like model, the electron density is described with the following double-exponential function:
\begin{equation}
	n_e(r, z) = n_0 \cdot \exp\left(-\frac{r}{r_0}\right) \cdot \exp\left(-\frac{|z|}{z_0}\right), \label{eq10}
\end{equation}
where \textit{r} and \textit{z} are the radial distance and the vertical height from the Galactic center in cylindrical coordinates, respectively; $n_0$ is the electron density at the Galactic center, $r_0$ is the scale length, and $z_0$ is the scale height. We generated 400 disk-like EM models based on Eq. (\ref{eq6}) with the following range of model parameters: $r_0$ from 1 kpc to 40 kpc with 1 kpc increments, and $z_0$ from 1 kpc to 10 kpc with 1 kpc increments. The position of the Sun was set to (\textit{r}, \textit{z})=(8.5, 0) kpc. The value of $n_0$ was automatically determined by minimizing the $\chi^2$ value. The best-fit parameters were $n_0=8.4_{-2}^{+3} \times 10^{-3}$ cm$^{-3}$, $r_0=20_{-4}^{+6}$ kpc, and $z_0=6_{-2}^{+3}$ kpc, with the 1-$\sigma$ confidence range for the minimum reduced $\chi^2$$\sim$2.9. The corresponding vertical and longitudinal profiles are also shown in Figure \ref{fig13}. Each of the black solid, blue dashed, and red dot-dashed lines correspond to the sub-regions shown in the figure. As can be seen in the figure, the best-fit disk-like model is in good agreement with the observed EM profiles. 

We would like to note that the estimated scale height of $z_0=6_{-2}^{+3}$ kpc is marginally comparable to the scale height of $3.6_{-0.8}^{+1.0}$ kpc of \ion{C}{4}, but is larger than the scale height of $2.6_{-0.5}^{+0.5}$ kpc of \ion{O}{6} by a factor of $\sim$2 obtained from the absorption line measurement \citep{Sav2009}.  \citet{Wak2012} found that the high ions are distributed in a layer with a scale height of $\sim$3 kpc. In fact, due to the local variation and irregularity of the actual gas distribution, it is difficult to describe the actual gas distribution with a simple disk-like model in which the scale height is constant. For example, \citet{Bow2008} derived different scale heights of 3.2 kpc and 4.6 kpc for the \ion{O}{6} at southern sky and northern sky, respectively, and other previous studies have presented slightly different scale heights \citep{Sav1997, Sav2000, Sav2003}.  Though the observed values show large variations, the scale heights of \ion{C}{4} and \ion{O}{6} seem to reside in the range of 1 to 10 kpc. 

Using the determined density distribution model, it is easy to estimate the total mass of the transition-temperature gas. Using the double-exponential density model, which was obtained by excluding the Galactic plane region of |\textit{b}| < 15\degr{}, the total mass based on the best-fit disk-like model is:
\begin{equation}
\begin{array}{l}
	M = 2 \int_{0}^{\infty} \int_{0}^{\infty} \mu m_p n_{tot} \cdot 2\pi r d{r} d{z} \\ 
	\;\;\;\;\; = 2 \mu m_p \cdot 1.92n_0 \cdot 2\pi z_0 r_0^2 \\ 
	\;\;\;\;\; = 3.2 \times 10^5 \cdot \left( \frac{n_0}{10^{-3} cm^{-3}} \right) \left( \frac{z_0}{kpc} \right) \left( \frac{r_0}{kpc} \right)^2 M_{\astrosun}, \label{eq11}
\end{array}
\end{equation}
where $\mu$ is the mean atomic weight of 0.6, $m_p$ is the proton mass, and $n_{tot}$ is the total gas density of $\sim$1.92$n_e$ for a fully ionized gas. The resulting total mass of the transition-temperature gas is $6.4_{-2.8}^{+5.2}\times10^9 M_{\astrosun}$. It is approximately three times as large as the total mass of 2$\times10^{9} M_{\astrosun}$ of the hot gas halo derived for the solar metallicity in \citet{Nak2018}. Using Eq. (12) of \citet{Nak2018}, the total mass of hot gas is estimated to be 6$\times10^{9} M_{\astrosun}$ for a lower metallicity of $Z\sim0.3Z_{\astrosun}$. If this is the case, then the total mass of the transition-temperature gas would be comparable to that of the hot gas. We note, however, that the derived mass of the transition-temperature gas is susceptible to the scale height and scale length and thus it has considerable uncertainty. The total mass of hot gas also sensitively depends on the assumed metallicity. However, it is clear, even after taking these uncertainties, that the total mass of the transition-temperature gas is still too small compared to the missing baryon mass of $\sim$2$\times10^{11} M_{\astrosun}$ \citep{Mil2015}.

\subsection{Volume-filling Fraction of the Transition-temperature Gas}

While various components of ISM have been observed from X-rays to radios for the entire Galaxy, the relative proportions between these components are still being debated \citep{Fer2001}. In this regard, information on the amount as well as the volume-filling fraction of the transition-temperature gas can provide further constraints on the configuration of the ISM, especially for the interstellar hot gas. As interstellar hot gas is expected to be supplied primarily by supernova explosions, the volume-filling fraction of the hot gas is closely related to the evolution model of supernova remnants. \citet{Cox1974} proposed a porosity factor \textit{q}, as defined by:
\begin{equation}
	q = S \int_{0}^{\infty} V_{SNR} d{t} = S V_{SNR} \tau, \label{eq12}
\end{equation}
where \textit{S} is the average supernova occurrence rate per unit volume, $V_{SNR}$ is the average volume of a supernova remnant at its final stage, and $\tau$ is the lifetime of an isolated supernova remnant. If \textit{q} is larger than unity, the supernova remnants overlap each other. The volume-filling fraction \textit{f} is defined as $f=1-e^{-q}$. With $S=9\times 10^{-14}$ pc$^{-3}$ yr$^{-1}$, $R_{SNR}=40$ pc, and $\tau=4\times 10^6$ yr; \citet{Cox1974} obtained $q\approx f\simeq 0.1$, and suggested that only a small fraction of the Galactic space is occupied by the hot gas which percolates through ISM. \citet{McK1977} recommended that \textit{q} > 2.9, which corresponds to \textit{f} > 0.95, implying that most of the Galactic space is filled with hot gas. On the contrary, \citet{Sla1993a}, with an improved simulation model and supernova parameters, suggested $q\approx f\simeq 0.18$. As can be seen, the volume-filling fraction of hot gas is highly dependent on the supernova explosion rate, which has large uncertainties. These methods of calculating the volume-filling fraction of hot gas mainly focus on the average hot gas distribution in the Galactic disk of |\textit{z}| < 300 pc. Meanwhile, despite the presence of a significant amount of hot gas in the Galactic halo, the volume-filling fraction of hot gas in the Galactic halo is still not well known.

\begin{figure*}[t]
	\begin{centering}
		\medskip{}
		\par\end{centering}
	\begin{centering}
		\includegraphics[clip,scale=0.7]{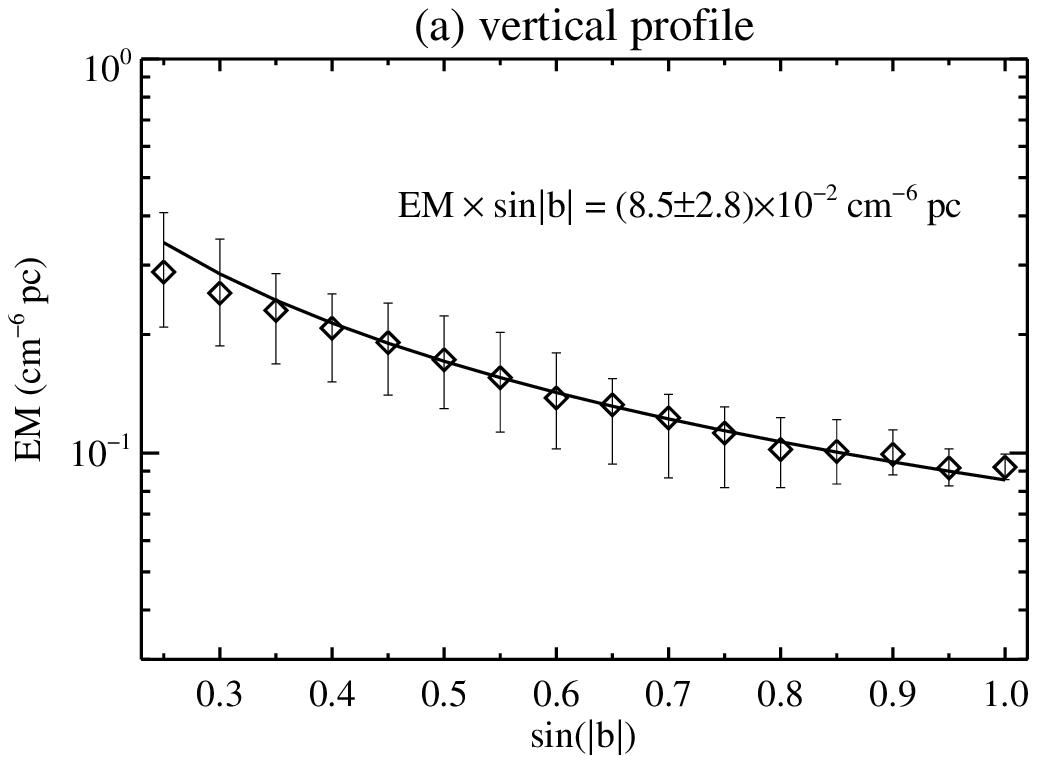}\
		\includegraphics[clip,scale=0.7]{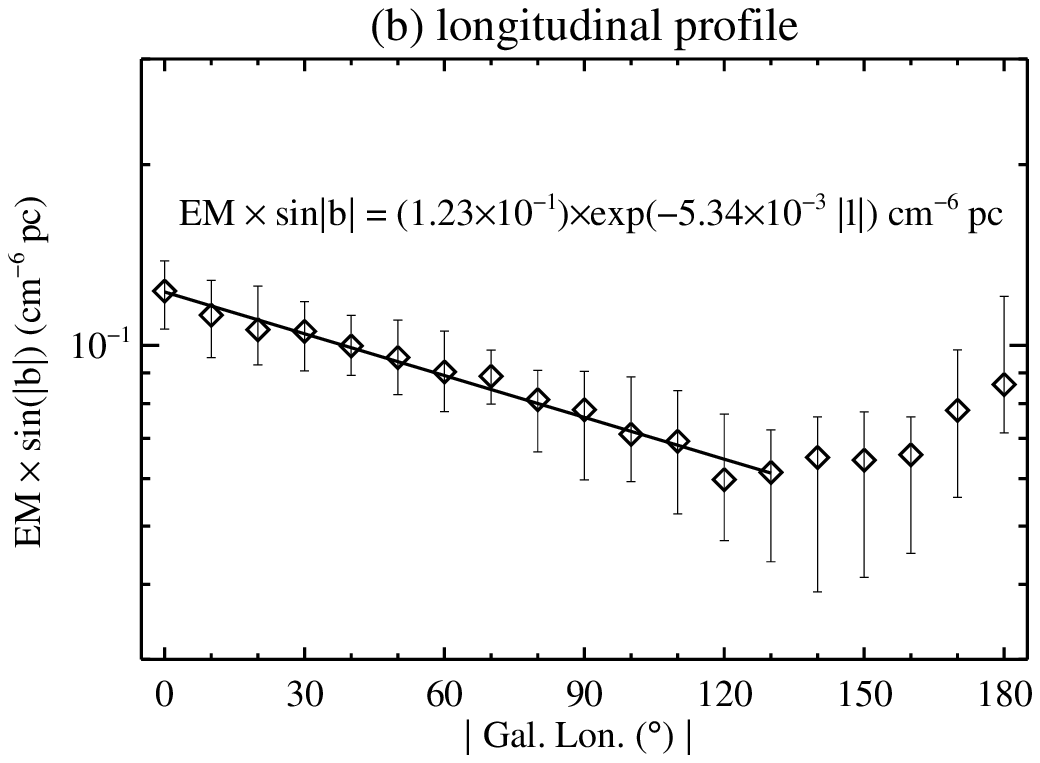}
		\par\end{centering}
	\begin{centering}
		\medskip{}
		\par\end{centering}
	\caption{\label{fig14}
		Average profiles of the emission measure with fitted models: (a) the vertical profile of the emission measure averaged over the Galactic longitudes at each Galactic latitude bin, and (b) the longitudinal profile averaged over the Galactic latitudes (|\textit{b}| > 15\degr{}) at each Galactic longitude bin. The black solid lines in (a) and (b) are the best-fit model of constant $EM\cdot \sin{|b|}$ for the vertical profile and the best-fit exponential function for $EM\cdot \sin{|b|}$ for the longitudinal profile, respectively. }
\end{figure*}

Now let us estimate the volume-filling fraction based on the present observation of high ions, though the distribution of the transition-temperature gas may be different from that of the hot gas. We can define the volume-filling fraction of the transition-temperature gas to be the ratio of the length of gas parcels, which the transition-temperature gas occupies, to the scale length along the line of sight. The length $\lambda$ can be obtained from:
\begin{equation}
	EM = \int_{0}^{\infty} n_e^2 d{x} \cong \left<n_e\right>^2\lambda, \;\; if \; \left<n_e^2\right> = \left<n_e\right>^2, \label{eq13}
\end{equation}
where the mean electron density $\left<n_e\right>=1.04\times 10^{-2}$ cm$^{-3}$ for a fully ionized gas, assuming a constant gas pressure of $P=3.8\times 10^{3}$ cm$^{-3}$ K \citep{Jen2011} and constant gas temperature of $T=1.9\times 10^{5}$ K (this work). For the scale length \textit{L} of EM, we assumed a simple plane parallel model and ignored the radial dependence in the previous disk-like model. The scale length simply becomes $L=0.5z_0\cdot \csc{|b|}$. Then, the volume-filling fraction \textit{f} of the transition-temperature gas is:
\begin{equation}
	f = \frac{\lambda}{L} = \frac{EM/\left<n_e\right>^2}{0.5z_0\cdot \csc{|b|}} = \frac{2EM\cdot \sin{|b|}}{z_0\cdot \left<n_e\right>^2}, \label{eq14}
\end{equation}
with $z_0=6$ kpc and $\left<n_e\right>=1.04\times 10^{-2}$ cm$^{-3}$, the volume-filling fraction \textit{f} is proportional to $EM\cdot \sin{|b|}$. Figure \ref{fig14}a shows the plot of the observed vertical profile of the EM with black diamonds, averaged in the radial direction, against $\sin{|b|}$. We fitted the profile of $EM\cdot \sin{|b|}$ with a constant value of (8.5$\pm$2.8)$\times10^{-2}$ cm$^{-6}$ pc, which yields the volume-filling fraction of $f=0.26\pm0.09$. To see the longitudinal dependence, we plotted $EM\cdot \sin{|b|}$, averaged in the vertical direction, along the Galactic longitude, as in Figure \ref{fig14}b. For the case of the \ion{C}{4} and \ion{O}{6} intensities in Figure \ref{fig6}b, $EM\cdot \sin{|b|}$ decreases until |\textit{l}|$\sim$130\degr{} and begins to increase gradually with increasing |\textit{l}|. The black solid line is the best-fit exponential function for the region of |\textit{l}| < 130\degr; $EM\cdot \sin{|b|}$ is reduced to $6\times 10^{-2}$ cm$^{-6}$ pc at \textit{l} = 130\degr{} from $1.2\times10^{-1}$ cm$^{-6}$ pc at \textit{l} = 0\degr. The corresponding volume-filling fraction is $f\sim0.18$ in the outer Galaxy (\textit{l} = 130\degr) and $f\sim0.37$ in the inner Galaxy (\textit{l} = 0\degr). The smaller volume-filling fraction of the outer Galaxy may indicate relatively weaker interactions between the hot and cold gases compared to the inner Galaxy.

\begin{figure}[t]
	\begin{centering}
		\medskip{}
		\par\end{centering}
	\begin{centering}
		\includegraphics[clip,scale=0.7]{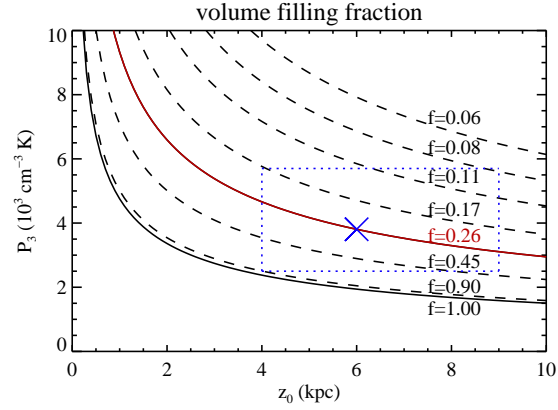}
		\par\end{centering}
	\begin{centering}
		\medskip{}
		\par\end{centering}
	\caption{\label{fig15}Volume-filling fraction as a function of the scale height ($z_{0}$) and gas pressure ($P_{3}$). Each contour line corresponds to the displayed volume-filling fraction. The blue cross indicates the best volume-filling fraction of $f=0.26$ for $z_{0}=6$ kpc and $P=3.8\times10^3$ cm$^{-3}$ K. The blue dotted box indicates the error range of $z_{0}$ and $P_{3}$.}
\end{figure}

We would like to note that the volume-filling fraction is sensitive to the scale height and gas pressure which are not actually constant. Figure \ref{fig15} shows the contour lines of the volume-filling fraction for the scale height ($z_0$, kpc) and gas pressure ($P_3$ with unit of $10^3$ cm$^{-3}$ K). The volume-filling fraction can vary from 0.39 to 0.17 for the scale height of $z_0$ = 4 kpc to 9 kpc, respectively, for the constant gas pressure of $P=3.8\times 10^{3}$ cm$^{-3}$ K assumed here. The volume-filling fraction even varies from 0.60 to 0.12 for the gas pressure of $P=2.5-5.7\times 10^{3}$ cm$^{-3}$ K \citep{Jen2011}, respectively, for the constant scale height of $z_0$ = 6 kpc. The ranges for $z_0$ and $P_3$ mentioned above are presented in the blue dotted box in Figure \ref{fig15}. Nevertheless, it is worth emphasizing the fact that the volume-filling fraction of \textit{f} = 0.26 is comparable to that of $f\sim0.25$ estimated from the 3-D supernova-driven ISM simulation for the Galactic disk (|\textit{z}| $\lesssim$ 250 pc) by \citet{deA2004}.

\section{Discussion}

\citet{Kor1998} pointed out that the Galactic morphologies of \ion{C}{4} and \ion{O}{6} differ significantly according to the hot gas generation models, indicating the importance of an all-sky survey of the FUV emission lines in discriminating the models. The spatial distribution of the FUV emission lines is expected to be plane-brightened in the isolated supernova-type model \citep{Cox1974}, fairly uniform in the three-phase model \citep{McK1977}, and highly pole-brightened in the Galactic fountain model \citep{Sha1976}. Whereas the intensity profiles of \ion{C}{4} and \ion{O}{6}, shown in Figure \ref{fig6}, are well-described by a simple plane-parallel model, but with a scale height much larger than the theoretical prediction of $\sim$300 pc in the supernova evolution model.

\begin{figure}[t]
	\begin{centering}
		\medskip{}
		\par\end{centering}
	\begin{centering}
		\includegraphics[clip,scale=0.9]{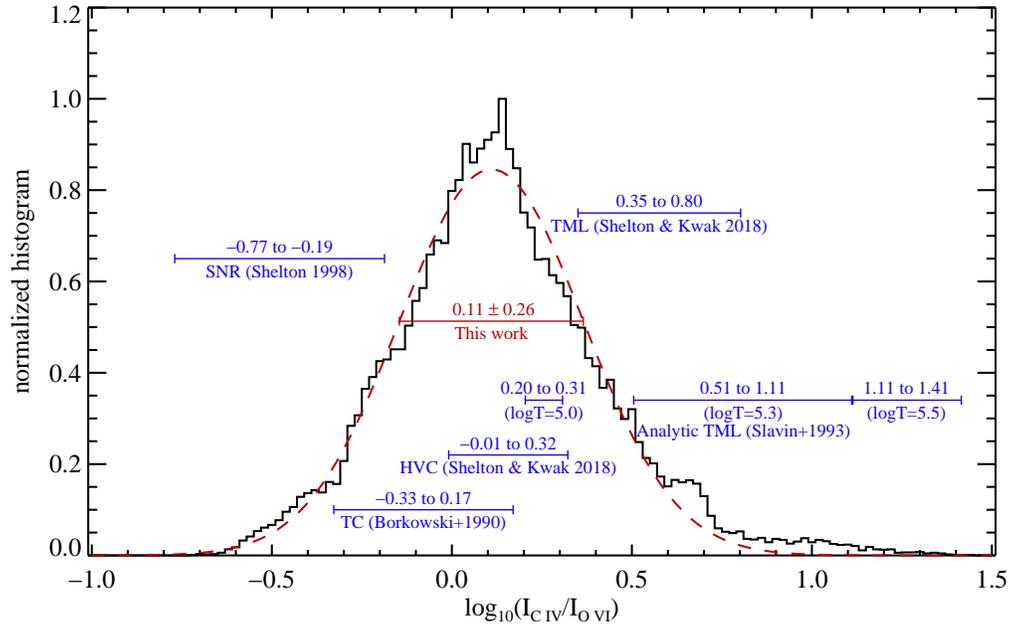}
		\par\end{centering}
	\begin{centering}
		\medskip{}
		\par\end{centering}
	\caption{\label{fig16} Normalized histogram of the line ratio of log(I(\ion{C}{4})/I(\ion{O}{6})) for the pixels with |\textit{b}| > 15\degr{}. Also over-plotted are the theoretical predictions based on various simulations : supernova remnant (SNR) by \citet{She1998}, thermal conduction (TC) by \citet{Bor1990}, high velocity clouds (HVC) by \citet{She2018}, analytic turbulent mixing layers (TML) by \citet{Sla1993b}, and TML by \citet{She2018}.}
\end{figure}

With the typical values of the parameters, such as the supernova explosion rate of $S=0.4\times10^{-13}$ pc$^{-3}$ yr$^{-1}$, gas density of $n_0=0.15$ cm$^{-3}$, gas pressure of $P=0.9\times10^4$ cm$^{-3}$ K, and their scale heights of $H_{SN}$ = 300 pc and $H_n \sim H_p \sim 340$ pc, respectively, we may estimate the \ion{C}{4} intensity (I(\ion{C}{4})) at high Galactic latitudes for the supernova evolution model using Eq. (14) of \citet{Shu1994}; The result is approximately 940 LU, which accounts for only $\sim$20\% of the I(\ion{C}{4})$\sim$5200 LU observed in the Galactic pole region of \textit{b} > 60\degr{} in the present study. \citet{Kor1998} also noted that the isolated supernova-type model may explain only $\sim$10\% of the observed \ion{C}{4} intensity. Furthermore, since I(\ion{C}{4}) is proportional to EM (or $n_e^2$) for a given gas temperature, the scale height of I(\ion{C}{4}) is estimated to be $\sim$3 kpc for the scale height $z_0=6_{-2}^{+3}$ kpc of $n_e$ estimated in the present study, which is larger than 300 pc of the supernova evolution model by a factor of 10. \citet{Shu1994} pointed out that the \ion{C}{4} intensity might not be accounted for solely by the supernova remnants near the Galactic disk and suggested that superbubbles and turbulent mixing layers (TML) at large distances above the Galactic plane would be the plausible candidates for additional \ion{C}{4} emission sources. In fact, the FUV line intensity, shown in Figure \ref{fig4}, is noticeably strong in the large bubbles close to the Sun, such as North polar spur or Orion-Eridanus. Although superbubbles far from the Sun ($\gtrsim$1 kpc) may not be well-distinguished in our observations, there is expected to be a large number of distant superbubbles, significantly contributing to the intensity and the scale height of \ion{C}{4}. However, it is unlikely that supernova remnants or superbubbles are the main production mechanism of high ions. According to a supernova remnant simulation by \citet{She1998}, the line ratio of log(I(\ion{C}{4})/I(\ion{O}{6})) ranges from -0.77 to -0.19 within 10 Myr while our observational result is log(I(\ion{C}{4})/I(\ion{O}{6})) = 0.11$\pm$0.26, as shown in Figure \ref{fig16}. Figure \ref{fig16} shows the normalized histogram of the line ratio of log(I(\ion{C}{4})/I(\ion{O}{6})) for the pixels with |\textit{b}| > 15\degr{}. Various theoretical predictions for the line ratio are also over-plotted in the figure for comparisons. The relatively higher observed line ratio indicates that the main mechanism originates from something different to a supernova remnant. Meanwhile, the predicted values of the line ratio of the TML models are generally larger than the observed values. The TML models by \citet{Sla1993b} predict log(I(\ion{C}{4})/I(\ion{O}{6})) = 0.20--1.41, and the TML models by \citet{She2018} predict log(I(\ion{C}{4})/I(\ion{O}{6})) = 0.35--0.80. The combined effects of TML and supernova remnants may explain the observed line ratio.

\citet{Wak2012} examined various theoretical predictions from the absorption study of high ions for 58 extragalactic targets observed with the FUSE and Space Telescope Imaging Spectrograph (STIS). By comparison of the column densities and ionic ratios of \ion{O}{6}, \ion{N}{5}, \ion{C}{4}, and \ion{Si}{4} to a series of theoretical predictions, they concluded that nonequilibrium ionization (NEI) radiative cooling is important in generating the transition temperature gas. They also suggested that the thick disk supernova and TML could play an important role in explaining the high ions in the Galactic corona. In contrast, thermal conduction (TC) and CIE were considered to be less important because they are insufficient to account for the observed \ion{Si}{4}. \citet{She2018} tested various models by comparing them with the ratios of the intensities of various emission lines observed by FIMS/SPEAR. They found that the closest models to observation include one of the TC front models of \citet{Bor1990}, the two TML models of \citet{Sla1993b}, and the HVC models of \citet{She2018}. Although their observations were limited to only four areas, their conclusion is consistent with our result shown in Figure \ref{fig16}. However, they also concluded that, as with \citet{Wak2012}’s results, no model can exactly reproduce the observations. Nonetheless, the emission line measurements of the high ions for the all-sky will certainly assist in the accurate validation of theoretical models. Recent hydrodynamic simulations on a Galactic scale may determine the main production mechanism of the high ions. For this purpose, future all-sky survey UV missions would be required to observe not only \ion{C}{4} and \ion{O}{6} but also other high ions, such as \ion{Si}{4} and \ion{N}{5}, with a higher spectral resolution and throughput.

\section{Summary}

In this study, we constructed and analyzed the maps of \ion{C}{4} and \ion{O}{6} emission lines, covering approximately 85\% and 77\% of the sky, respectively, using the archival data observed by FIMS/SPEAR. We employed an adoptive smoothing method to secure a sufficiently significant signal-to-noise ratio. The average FWHMs of the \ion{C}{4} and \ion{O}{6} maps are 3.2\degr{} $\pm$ 1.8\degr{} and 4.5\degr{} $\pm$ 3.1\degr{}, respectively. The \ion{C}{4} and \ion{O}{6} intensities were corrected for the extinction under the assumption of the uniformly mixed interstellar dust and sources.

The resulting \ion{C}{4} and \ion{O}{6} line maps are described well by a simple plane-parallel model or a double exponential model. The median values of $I\times\sin{|b|}$ for the \ion{C}{4} and \ion{O}{6} lines are approximately 4500 LU and 3400 LU, respectively. We further estimated the gas temperature and the emission measure assuming isothermal plasma. Despite the wide variation of $I_{O VI}/I_{C IV}$, the gas temperature is well-defined within the narrow range of $(1.89 \pm 0.06) \times 10^5$ K. The gas temperature was found to be generally uniform regardless of the Galactic latitude, but appears higher in the region of |\textit{l}| > 90\degr{} than in the region of |\textit{l}| < 90\degr{}. This may indicate that the hot gases originating from the supernova explosions in the disk escape more easily toward the halo in the outer Galaxy region, producing higher temperature gas there, than in the inner Galaxy region. The emission measure of the transition-temperature gas is well-described by a disk-like model, of which the best-fit parameters are $n_0=8.4_{-2}^{+3} \times 10^{-3}$ cm$^{-3}$, $r_0=20_{-4}^{+6}$ kpc, and $z_0=6_{-2}^{+3}$ kpc. The total mass of the transition-temperature gas estimated from these best-fit parameters is $6.4_{-2.8}^{+5.2}\times10^9 M_{\astrosun}$, corresponding to approximately 3\% of the missing baryon mass of $\sim$2$\times10^{11} M_{\astrosun}$ of which a significant fraction is believed to be in the extended halo.

The volume-filling fraction of the transition-temperature gas is estimated to be $f=0.26\pm0.09$, and it varies from $f\sim0.37$ in the inner Galaxy to $f\sim0.18$ in the outer Galaxy. The smaller volume-filling fraction of the outer Galaxy may indicate relatively weaker interactions between the hot and cold gases compared to the inner Galaxy.

The spatial distribution of the FUV emission lines is not explained by traditional models, such as the isolated supernova-type model by \citet{Cox1974} and the three-phase model by \citet{McK1977}. As the intensity ratio of \ion{C}{4} and \ion{O}{6} is between that of the supernova remnant model and that of the TML model, the combined effects of supernova remnants and TML may explain the observed line ratio. Our result is consistent with that of \citet{She2018}, of which the thermal conduction front model of \citet{Bor1990}, the two TML models of \citet{Sla1993b}, and their HVC models are most consistent with these observations.

\acknowledgements{This research was supported by the Korea Astronomy and Space Science Institute under the R\&D program supervised by the Ministry of Science, ICT, and Future Planning of Korea. This research was also supported by the Basic Science Research Program (2017R1D1A1B03031842) through the National Research Foundation (NRF) funded by the Ministry of Education of Korea. This work was also supported by the NRF grant of Korea, with grant number NRF-2014M1A3A3A02034746. K.-I. Seon was supported by the NRF grant funded by the Korea government (MSIP; No. 2017R1A2B4008291)}

\pagebreak{}

\begin{continuedfigure}[tp]
	\begin{centering}
		\medskip{}
		\par\end{centering}
	\begin{centering}
		\includegraphics[clip,scale=0.6,angle=90]{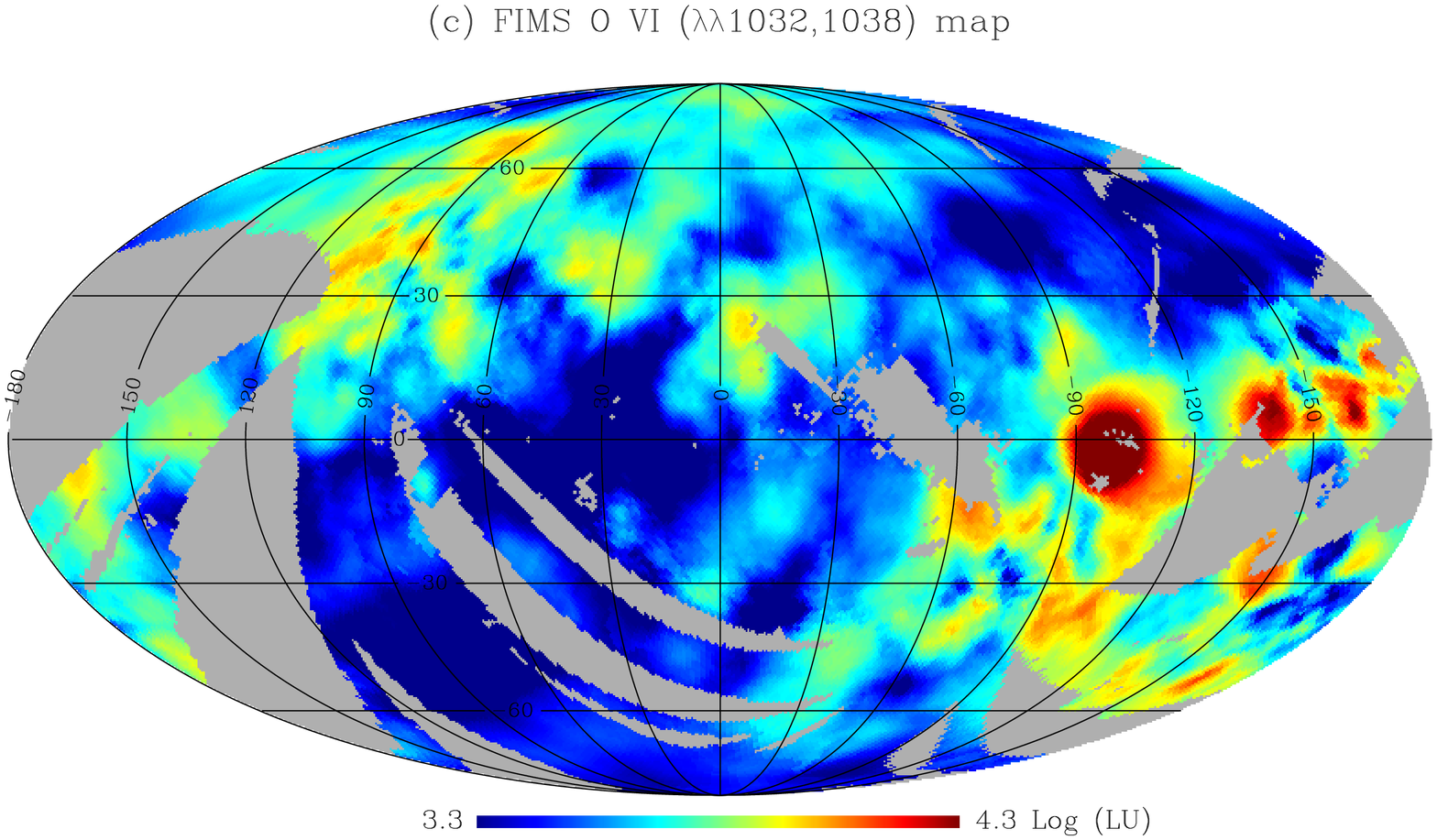}
		\par\end{centering}
	\begin{centering}
		\includegraphics[clip,scale=0.6,angle=90]{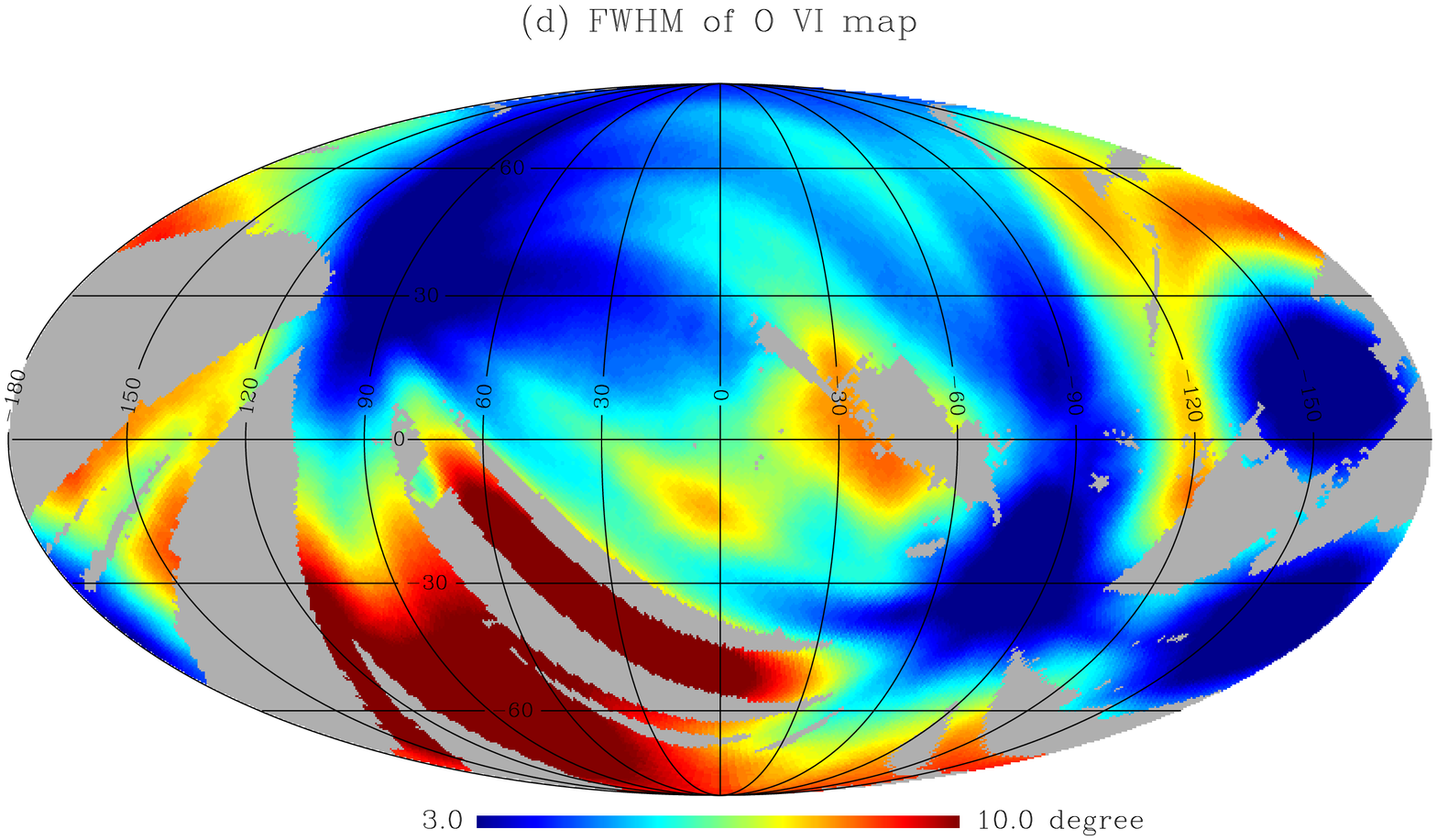}
		\par\end{centering}
	\begin{centering}
		\medskip{}
		\par\end{centering}
	\caption{Continued.}
\end{continuedfigure}

\end{document}